\documentclass[numberedappendix,twocolappendix,iop]{emulateapj}
\usepackage{color}
\def\bmath#1{\mbox{\boldmath$#1$}}
\def\revise{}
\slugcomment{Accepted for publication in ApJ}

\shorttitle{} 
\shortauthors{Matsumoto et al.}

\begin{document}

\title{Star formation in turbulent molecular clouds with colliding flow}

\author{
  Tomoaki Matsumoto\altaffilmark{1}, 
  Kazuhito Dobashi\altaffilmark{2}, and 
  Tomomi Shimoikura\altaffilmark{2}
}

\altaffiltext{1}{Faculty of Humanity and Environment, Hosei University, Fujimi, Chiyoda-ku, Tokyo 102-8160, Japan}
\email{matsu@hosei.ac.jp}
\altaffiltext{2}{Department of Astronomy and Earth Sciences, Tokyo Gakugei University, Koganei, Tokyo 184-8501, Japan}

\begin{abstract}
Using self-gravitational
hydrodynamical numerical simulations, we investigated the evolution of high-density turbulent molecular clouds
swept by a colliding flow.  
The interaction of shock waves due to turbulence produces networks of
thin filamentary clouds with a sub-parsec width. 
The colliding flow accumulates the
filamentary clouds into a sheet cloud and promotes active star
formation for initially high-density clouds. 
Clouds with a colliding flow exhibit a finer filamentary network 
than clouds without a colliding flow.
The probability distribution functions (PDFs) for the density and column
density can be fitted by lognormal
functions for clouds without colliding flow.  When the initial
turbulence is weak, the column density PDF has a power-law wing at high column densities.
The colliding flow considerably deforms the PDF,
such that the PDF exhibits a double peak.
The stellar mass distributions reproduced here are consistent with 
the classical initial mass function with a power-law index of $-1.35$ when the initial
clouds have a high density.
The distribution of stellar velocities agrees with the gas
velocity distribution, which can be fitted by Gaussian functions
for clouds without colliding flow.  For clouds with colliding
flow, the velocity dispersion of gas tends to be larger than the
stellar velocity dispersion. 
The signatures of colliding flows and turbulence appear in channel
maps reconstructed from the simulation data.  Clouds without
colliding flow exhibit a cloud-scale velocity shear due to the
turbulence.  In contrast, clouds with colliding flow show
a prominent anti-correlated distribution of thin filaments between
the different velocity channels, suggesting collisions between
the filamentary clouds.
\end{abstract}


\section{Introduction}

Turbulence of the interstellar medium plays an important role in star
formation \citep{Elmegreen04,Scalo04,McKee07}. 
It is widely accepted that 
turbulence controls the fragmentation of molecular clouds. 
When the fragments are gravitationally unstable,
dense fragments caused by strong shocks
undergo collapse to form stars \citep[e.g.,][]{Klessen00a}.
Consequently, characteristics of star formation, such as star-formation
rates and star-formation efficiency, are presumed to be affected by 
turbulence \citep{Krumholz05,Padoan11,Hennebelle11}.

In addition to turbulence, molecular clouds are affected
by external effects.
It has been thought that 
H\,{\sc ii} regions and stellar winds from OB stars trigger star
formation in associated molecular clouds \citep{Elmegreen77}.
For example, supernova remnants interact with the interstellar medium
\citep{McKee77}. Observations suggest that 
supernova remnants also trigger star formation
\citep[e.g.,][]{Koo08,Gouliermis08}, although 
this manner of formation is still under debate \citep{Desai10}.

Theoretically, \citet{Hartmann01} proposed a scenario of
dynamical star formation, where large-scale flow in the interstellar medium accumulates gas
to promote rapid star formation.
\citet{Vazquez07}, \citet{Banerjee09}, and
\citet{Gomez14}
examined the formation of giant molecular clouds
and stars within them by considering the collision of a warm neutral
medium. 
\revise{
\citet{Gong11} and \citet{Chen14} considered cloud core formation in
turbulent clouds with supersonic converging flows, by using
hydrodynamical and ambipolar-diffusion magnetohydrodynamics (MHD) simulations.
}
These theoretical studies indicate that both the external flow
and the turbulence
are important factors influencing \revise{cloud core formation} and star formation.

Recently, \citet{Dobashi14} carried out molecular line observations
of the massive dense cloud L1004E, which is located
in the Cyg~OB~7 giant molecular cloud.  
The characteristic features of L1004E
are a high mass of $\sim 10^4\,M_\sun$ within a small region $\sim (5~\mathrm{pc})^2$, 
a low temperature of $\sim 9$~K, 
and a broad line width of $\sim 2\,\mathrm{km\,s}^{-1}$
corresponding to Mach 10 turbulence. 
A small number of protostar candidates associated with the cloud
suggests that the cloud is probably in a stage prior to active star
formation or has just begun forming stars. 
Moreover, their observations suggest a dynamical process inside
L1004E, where several filamentary clouds can be identified as internal structures
that interact with each other by collisions. 
The interaction may induce star formation.
They also suggested that the interaction between filaments is
caused by compression due to an external effect, e.g., the shock of nearby
supernova remnants. 

Recent observations indicate that filaments are the basic unit of
structure in molecular clouds.
Observations by the {\it Herschel} survey revealed
networks of parsec-scale filaments in many clouds
\citep[e.g.,][]{Andre10,Menshchikov10,Arzoumanian11}. 
\citet{Hacar13} showed that filamentary cloud L1495/B213 in the Taurus
region has a substructure containing many filaments.
\citet{Nakamura14} suggested that the collisions of filaments may have triggered cluster formation.

Motivated by the observations of \citet{Dobashi14} 
and taking into account compression 
on a cloud scale, 
we performed numerical simulations of self-gravitational turbulent molecular
clouds.  Although
self-gravitational turbulent gas has been investigated by many authors
\citep[e.g.][]{Klessen00a,Ostriker01,Heitsch01,Federrath08,Nakamura08,Offner08,Padoan11},
simulations are still limited for clouds with strong
turbulence and low temperature, such as L1004E \citep{Padoan11}.
Moreover, the influence of cloud-scale compression on star formation
in these clouds is poorly understood.

This paper is organized as follows.  In Section~\ref{sec:models} and
Section~\ref{sec:methods}, the model and simulation methods are
presented.  The results of the simulations are shown in
Section~\ref{sec:results}, and they are discussed in
Section~\ref{sec:discussion}.  Finally, conclusions of this paper are
given in Section~\ref{sec:summary}.

\section{Models of Turbulent Clouds}
\label{sec:models}

As the initial condition, we consider clouds with a constant density and
turbulent velocity. The clouds are swept by head-on colliding flows.
The initial models are classified according to the initial density $\rho_0$, 
the initial Mach number of turbulence ${\cal M}_t$, 
and the Mach number for a colliding flow ${\cal M}_f$.
The model parameters are summarized in Table~\ref{table:model-parameters}.

\begin{deluxetable*}{lcccrrcrc}
\tabletypesize{\scriptsize}
\tablecaption{Model parameters \label{table:model-parameters}}
\tablehead{
\colhead{Model} &
\colhead{$\rho_0$} &
\colhead{$n_0$} &
\colhead{$M_\mathrm{tot}$} &
\colhead{${\cal M}_t$} &
\colhead{${\cal M}_f$} &
\colhead{$t_\mathrm{ff}/t_\mathrm{cross}$} &
\colhead{$N_\star$\tablenotemark{a}} &
\colhead{$t_{10M_\sun}$\tablenotemark{b}}
\\
\colhead{} &
\colhead{($10^{-21} \mathrm{g}\,\mathrm{cm}^{-3}$)} &
\colhead{($10^3 \mathrm{cm}^{-3}$)} &
\colhead{($10^3 M_\sun$)} &
\colhead{} &
\colhead{} &
\colhead{} &
\colhead{} &
\colhead{($10^6$ yr)}
}
\startdata
HT3F0   & 5.42 & 1.42\phn  & 10.0 &  3 & 0 & 0.11 & 135 (13) & 1.43\\
HT10F0  & 5.42 & 1.42\phn  & 10.0 & 10 & 0 & 0.35 & 140 (12) & 0.87\\
HT30F0  & 5.42 & 1.42\phn  & 10.0 & 30 & 0 & 1.06 & 265 (19) & 0.76\\
HT10F3  & 5.42 & 1.42\phn  & 10.0 & 10 & 3 & 0.36 & 260 (15) & 0.87\\
HT10F10 & 5.42 & 1.42\phn  & 10.0 & 10 & 10& 0.43 & 142 (11) & 0.56\\
HT10F30 & 5.42 & 1.42\phn  & 10.0 & 10 & 30& 0.83 & 915 (54) & 0.50\\
LT3F0   & 1.00 & 0.262 & 1.85 &  3 & 0 & 0.25 & 14   \phn(0) & 3.16\\
LT10F0  & 1.00 & 0.262 & 1.85 & 10 & 0 & 0.82 & 9    \phn(0) & 2.51\\
LT30F0  & 1.00 & 0.262 & 1.85 & 30 & 0 & 2.46 & 3    \phn(1) & 2.21\\
LT10F3  & 1.00 & 0.262 & 1.85 & 10 & 3 & 0.84 & 25   \phn(0) & 2.08\\
LT10F10 & 1.00 & 0.262 & 1.85 & 10 & 10& 1.00 & 25   \phn(0) & 1.55\\
LT10F30 & 1.00 & 0.262 & 1.85 & 10 & 30& 1.92 & 55  (34) & 1.27\\
HT10F0wog\tablenotemark{c} & 5.42 & 1.42\phn & 10.0 & 10 & 0 &\nodata & \nodata &\nodata \\
HT10F10wog\tablenotemark{c}& 5.42 & 1.42\phn & 10.0 & 10 & 10&\nodata & \nodata &\nodata
\enddata
\tablenotetext{a}{Number of sink particle formed by the stage where the maximum mass of sink particles reaches $10M_\sun$.
Numbers in the parenthesis indicate the number of sink particles ejected from computational domains.}
\tablenotetext{b}{Time when the maximum mass of sink particles reaches $10M_\sun$.}
\tablenotetext{c}{Models without self-gravity.}
\end{deluxetable*}

The constant initial density is assumed to be $\rho_0 = 5.42\times 10^{-21}\,\mathrm{g}\,\mathrm{cm}^{-3}$ 
(corresponding to the number density $n_0 = 1.42 \times 10^3\,\mathrm{cm}^{-3}$)
for high-density models, and 
$\rho_0 = 10^{-21} \,\mathrm{g}\,\mathrm{cm}^{-3}$ 
($n_0 = 2.62\times 10^2 \,\mathrm{cm}^{-3}$)
for low-density models.
The computational domain is a cubic box with sides $L=5~\mathrm{pc}$,
and the total masses of gas are therefore
$M_\mathrm{tot} = 1.00 \times 10^4 \, M_\odot$ for the high-density models 
and 
$M_\mathrm{tot} = 1.85 \times 10^3\, M_\odot$ for the low-density models.
The mass of the high-density models mimics
the high-mass core L1004E in the Cyg~OB~7 molecular cloud \citep{Dobashi14}.

The gas is assumed to be isothermal with a temperature of 10~K, and 
the corresponding sound speed is $c_s = 0.19\,\mathrm{km}\,\mathrm{s}^{-1}$.
This temperature is in agreement with that in L1004E, 
which is estimated as $8~\mathrm{K} \lesssim T \lesssim 12~\mathrm{K}$ with a mean value
of $\sim 9$~K \citep{Dobashi14}. 
The initial Jeans length is defined by $\lambda_J = (\pi c_s^2/ G \rho_0)^{1/2}$,
and it is evaluated as 0.57~pc and 1.33~pc
for the high-density and low-density models, respectively.  
The ratio of the size of the computational domains
to the Jeans length is $L/\lambda_J = 8.7$ and 3.7
for the high-density and low-density models, respectively.  
Therefore, the total masses in units of the Jeans mass for the high-density models
and low-density models are 
$M_\mathrm{tot}/M_J = 6.59\times 10^2$ and $ 5.23 \times 10^1 $, respectively,
where the Jeans mass is defined by $M_J = \rho_0 \lambda_J^3$.
The freefall times are 
$t_\mathrm{ff} =9.05 \times 10^5$~yr and 
$2.11 \times 10^6$~yr, respectively, for the high-density and low-density models,
where the freefall time is defined by $t_\mathrm{ff} = (3\pi/32 G\rho_0)^{1/2}$.
\revise{The magnetic field is ignored in this paper for simplicity.}

A turbulent velocity is imposed on the initial stage and 
decay of the turbulence follows, under the consideration of no driving force for
turbulence \citep[see for detail][]{Matsumoto11a}.
The initial velocity field is solenoidal with a power
spectrum of $P(k) \propto k^{-4}$, generated according to
\citet{Dubinski95}, where $k$ is the wavenumber.  This power spectrum
results in a velocity dispersion of $\sigma_v(\lambda) \propto
\lambda^{1/2}$, where $\lambda$ is the scale length, 
which is in agreement with the Larson scaling relations
\citep{Larson81}.
Note that the kinetic energy of the turbulence is given by $E_K = \int P(k) 4
\pi k^2 dk$ in our definition.
An alternative definition of the power spectrum is $E(k) = 4\pi k^2 P(k)$,
and the kinetic energy is given by $E_K = \int E(k)dk$, yielding 
the relationship $E(k) \propto k^{-2}$.
The models are constructed by changing the root mean square (rms)
Mach number of the initial velocity field in the range 
${\cal M}_t = 3 - 30$. 
The rms Mach number is defined by
\begin{equation}
{\cal M}_t = 
\left[\frac{1}{c_s^2 L^3 } \int_V \left| \bmath{v}_t^2 \right| dV \right]^{1/2},
\end{equation}
where $\bmath{v}_t$ denotes the turbulent velocity and $\int_V dV$
denotes the volume integration over the computational domain.  
We utilize a common velocity pattern as the template
for producing the initial turbulence among all models.

To investigate the effects of a colliding gas flow, 
a large-scale sinusoidal flow is also imposed on the initial stage as
\begin{equation}
\bmath{v}_f = \left(
\begin{array}{c}
\displaystyle
{\cal M}_f c_s \sin \frac{2 \pi x }{ L } \\
0 \\
0
\end{array}
\right),
\end{equation}
where ${\cal M}_f$ denotes the amplitude of the flow in units of
Mach number, and it changes
in the range of ${\cal M}_f= 0 - 30$ for constructing the models.
Due to this flow, turbulent gas converges to the $x=0$ plane. 
This flow mimics the effects of the external environment, e.g., 
shock fronts of supernova remnants, H\,{\sc ii} regions, or stellar winds from
nearby OB stars. 

By using the two Mach numbers ${\cal M}_t$ and ${\cal M}_f$, 
the effective Mach number for the flows is defined by ${\cal
M}_\mathrm{eff} = ({\cal M}_t^2 + {\cal M}^2_f/2)^{1/2}$.  The factor
$1/2$ for ${\cal M}_f^2$
comes from the average of the sinusoidal flow $\bmath{v}_f$ over
the computational domain.
The crossing time is then evaluated as $t_\mathrm{cross}
= L/({\cal M}_\mathrm{eff} c_s) = 2.57 \times 10^6 \left({\cal
  M}_\mathrm{eff}/10\right)^{-1}\,\mathrm{yr}$.

As summarized in Table~\ref{table:model-parameters}, we call these models 
HT10F0, LT3F0, etc. 
The model names are constructed as follows. The 1st
character, ``H'' or ``L'', corresponds to the high-density
or low-density models, respectively. The digits following ``T'', which is the 
turbulence, 
denote the initial Mach number for the turbulence, ${\cal  M}_t$. 
The digits following ``F'', which is the flow,  denote the initial
Mach number for the colliding flow, ${\cal M}_f$.

We also calculated models HT10F0
and HT10F10 without considering self-gravity. These models are referred to as
``HT10F0wog'' and ``HT10F0wog'', and are calculated 
for the long time of $t \simeq 2.5 \times 10^7\,\mathrm{yr} \simeq 10 t_\mathrm{cross}$.
The turbulence decays, reducing the velocity dispersion
to $\simeq 7.9\%$ and 6.6\% of the initial value for models
HT10F0wog and HT10F10wog, respectively.

Note that the models presented here can be scaled to different
sizes because the gas is assumed to be isothermal.
The size of the problem is specified by the non-dimensional parameter
$L/\lambda_J$, or
equivalently $M_\mathrm{tot}/M_J$, or $t_\mathrm{ff} / t_\mathrm{sc} $, 
where $t_\mathrm{sc} (= L/c_s = 2.6\times 10 ^7\,\mathrm{yr)}$ denotes the sound crossing time.

\section{Numerical Methods}
\label{sec:methods}
We calculated the evolution of a cloud by 
the numerical simulation code, SFUMATO \citep{Matsumoto07}. 
The total variation diminishing (TVD) cell-centered scheme is adopted
as the hydrodynamical solver.
The hydrodynamical solver achieves second-order accuracy in space and time.
The self-gravity is solved by the multigrid method, which exhibits 
spatial second-order accuracy. 
The numerical fluxes are 
conserved by using the refluxing procedure in both the hydrodynamics and the self-gravity solvers.
The periodic boundary condition is imposed for the hydrodynamics and the self-gravity.

SFUMATO utilizes a block-structured adaptive mesh refinement (AMR) technique.
The computational domain of $(5~\mathrm{pc})^3$ is resolved by a
uniform grid having $512^3$ cells for the high-density models
and $256^3$ cells for the low-density models at the initial stage.
The initial resolutions are therefore
$\Delta x = 1.0\times 10^{-2}\,\mathrm{pc}$ for
the high-density models and
$\Delta x = 2.0\times 10^{-2}\,\mathrm{pc}$ for
the low-density models.
The Jeans condition is employed as a refinement criterion.
Blocks are refined when the Jeans length is shorter than 8 times the cell width:
$(\pi c_s^2/G\rho)^{1/2} < 8 \Delta x $ \citep[see][]{Truelove97}.
The finest resolution is set at $\Delta x_\mathrm{min} =
5.0\times10^{-3}\,\mathrm{pc}$ for both the high-density and
low-density models, and the effective mesh size is therefore $1024^3$.  
This mesh size is the same as that of the recent
high-resolution simulations \citep[e.g.,][]{Federrath10}.

Sink particles are implemented to reproduce
star formation in the model clouds. 
Sink particles are Lagrangian particles moving on the numerical
grid and interacting with the gas through gravity and accretion. 
The threshold density for particle creation is set at
$\rho_\mathrm{sink} = 1.17\times 10^{-18}\,
\mathrm{g}\,\mathrm{cm}^{-3}$ (corresponding number density is
$3.07\times 10^{5}\,\mathrm{cm}^{-3}$).
Fragmentation during gravitational collapse in a very high density regime ($\rho >
\rho_\mathrm{sink}$) does not occur, and the number of sink
particles exhibited in the simulations is therefore the lower limit of
the number of stars produced in the clouds.
Sink particles accrete gas located within a radius $ r_\mathrm{sink} = 1.95\times 10^{-2}\,\mathrm{pc}$, 
which is equal to $4\Delta x_\mathrm{min}$ as well as
half of the Jeans length for $\rho_\mathrm{sink}$.
The details of the method for sink particles are shown in Appendix~\ref{sec:sink_particle}.

We followed the evolution of the clouds until the stage in which the
maximum mass of sink particles exceeds $10\, M_\sun$.
Further evolution is beyond the scope of this paper because no
feedback from star formation is considered in the simulations.

\section{Results}
\label{sec:results}
\subsection{Fiducial model with high density HT10F0}
\begin{figure*}
\epsscale{1.}
\plotone{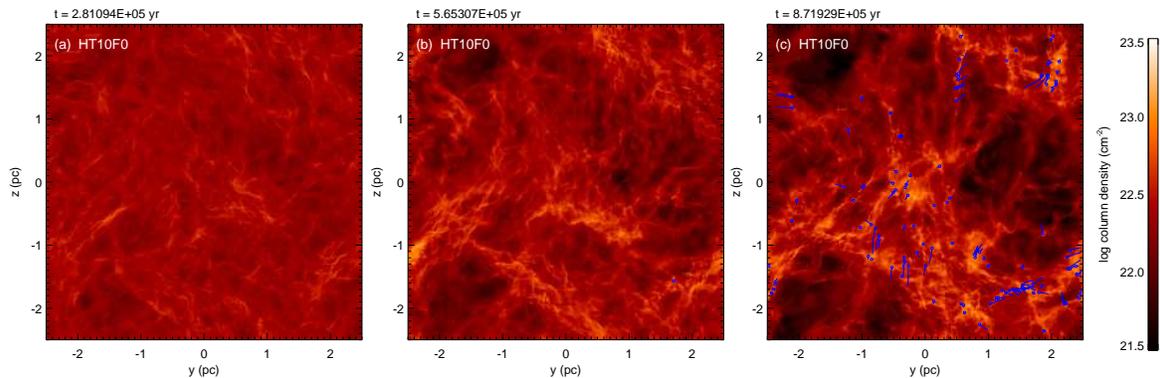}
\figcaption[]{ 
Evolution of model HT10F0 at three representative stages:
(a) the early stage, (b) the stage of the first star formation, and 
(c) the stage of $M_\mathrm{\star, max} = 10M_\sun$.
The color scale shows the column density in the $x$-direction.
The blue circles and their associated lines show the positions of sink particles
and their loci, respectively.
\label{colden_HT10F0.eps}
}
\end{figure*}

The evolution of the high-density model with ${\cal M}_t = 10$ and
${\cal M}_f =0$ is described here as a fiducial model without the
colliding flow.
Figure~\ref{colden_HT10F0.eps} shows the evolution of model HT10F0 by
the distributions of column density and sink particles. 
In the early phase, turbulence produces many shock waves, 
as shown in Figure~\ref{colden_HT10F0.eps}(a). 
The density, which increases by a factor of $\sim 10^2$ due to shock compression,
indicates that the density contrast of the isothermal shock is equal to 
the square of the Mach number.  The first star forms at 
$t = 5.65\times 10^5\,\mathrm{yr}$,
which corresponds $0.22 t_\mathrm{cross}$ and $0.62 t_\mathrm{ff}$,
as shown in Figure~\ref{colden_HT10F0.eps}(b). 
Sink particle formation earlier than the freefall time is
the result of the density increase due to the shock waves.
At this stage, models HT10F0 and HT10F0wog show similar density 
distributions, which indicates that turbulence mainly controls the
structure formation.
Figure~\ref{colden_HT10F0.eps}(c) shows the cloud at the stage where
the maximum mass of sink particles reaches $10\,M_\sun$ (hereafter
referred to as ``the stage of $M_\mathrm{\star, max} = 10\,M_\sun$'').
By this stage, the density contrast increases more than 
that shown in Figure~\ref{colden_HT10F0.eps}(b), and it is
slightly higher than that of model HT10F0wog because of self-gravity.
Sink particles are created in clusters along the dense filaments.

\subsection{Dependence on the initial strength of the turbulence}
\begin{figure*}
\epsscale{1.}
\plotone{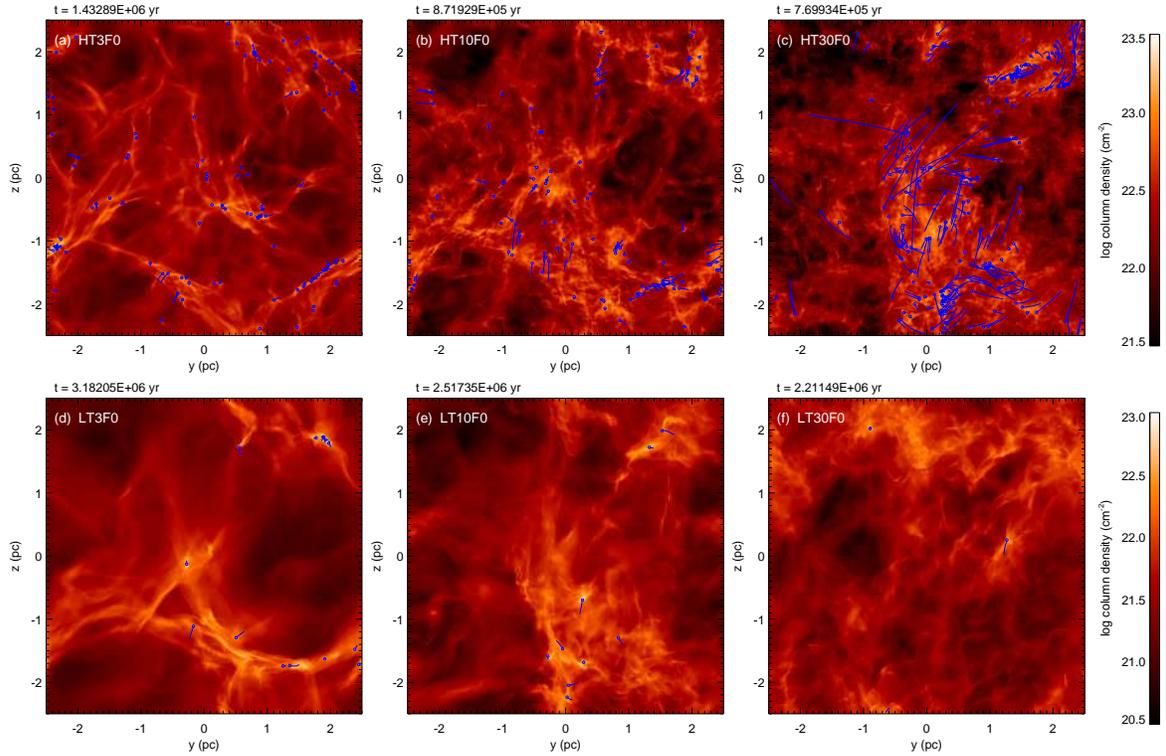}
\figcaption[]{ 
Comparison of clouds in models with different strengths of
initial turbulence and without the colliding flow (${\cal M}_f =0$)
at the stages of $M_\mathrm{\star, max} = 10M_\sun$.
The upper and lower figures show the high-density models and the low-density
models, respectively. 
The left, middle, and right figures show the models with 
${\cal M}_t = 3, 10,$ and 30, respectively.
The color scale shows the column density in the $x$-direction.
The blue circles and their associated lines show the positions of the sink particles
and their loci, respectively.
\label{colden_M.eps}
}
\end{figure*}

Figure~\ref{colden_M.eps} shows the dependence of the 
high-density models (upper panels) and 
low-density models (lower panels) without colliding flow 
(${\cal M}_f=0$) on the initial strengths of the
turbulence.

The stronger turbulent models produce a fine structure of gas,
whereas the weak turbulent models show parsec-scale filaments
irrespective of the initial densities.
Model HT30F0 exhibits small structures disturbed by strong
turbulence, whereas model HT3F0 exhibits a network with long filaments.
Similarly, model LT3F0 shows smoother filaments than does LT30F0.

For the high-density models, 
a model with stronger turbulence
produces more sink particles in a shorter time (see also Table~\ref{table:model-parameters}). 
This is attributed to strong compression due to a large amount of
turbulence, as shown in the previous work of \citet{Klessen00a}.
The long loci of sink particles in the more
turbulent model, HT30F0, indicate the high speeds of the sink particles.
These sink particles move at velocities roughly similar to 
those of the gas where they formed.
Details of the velocity distribution for sink particles are shown in
Section~\ref{sec:velocityDistribution}.

The low-density models produce much smaller numbers of sink
particles compared to the high-density models (see also Table~\ref{table:model-parameters}).
This indicates that the high-density models contain
larger masses than the low-density models.
\revise{
Contrary to the high-density models, 
a model with stronger turbulence produces fewer sink particles
by the stage of $M_\mathrm{\star, max} = 10\,M_\sun$, as
shown in the lower panels of Figure~\ref{colden_M.eps} and Table~\ref{table:model-parameters}. 
  However, at a given time, a model with stronger turbulence produces
  more sink particles, showing that the turbulence promotes
  star formation even for the low-density models. 
}

\subsection{Model with colliding flow HT10F10}
\label{sec:Model_with_colliding_flow_HT10F10}
\begin{figure*}
\epsscale{1.}
\plotone{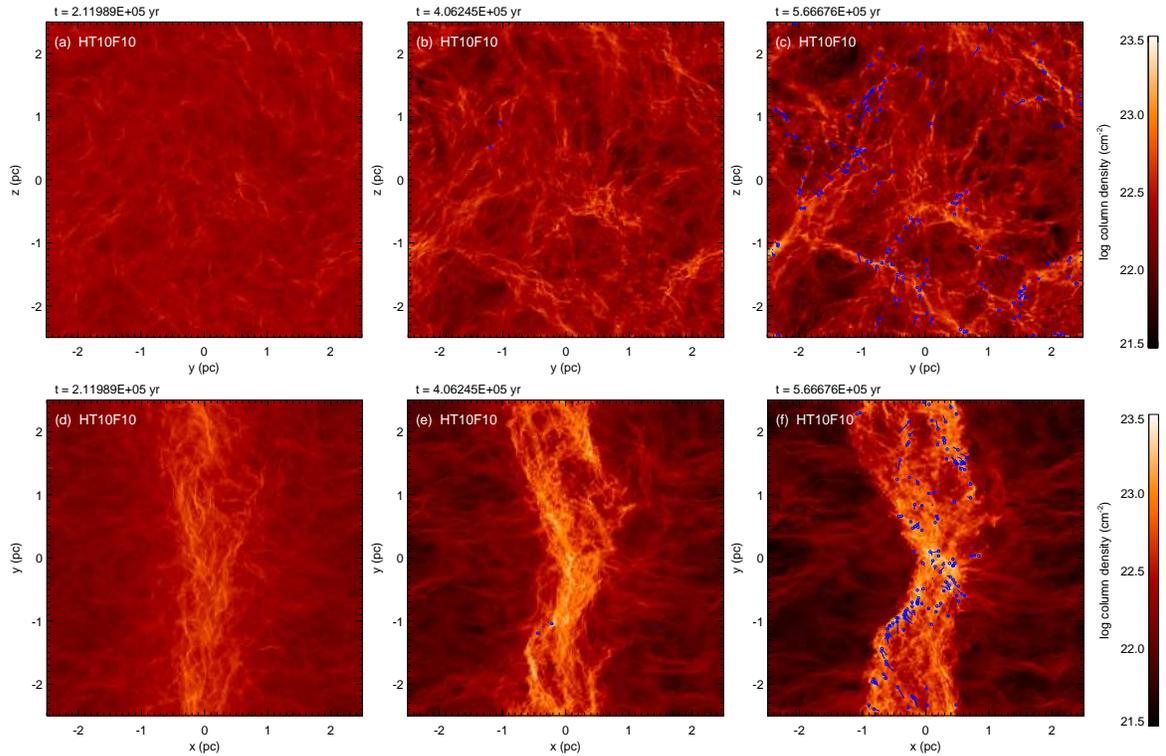}
\figcaption[]{ 
Evolution of model HT10F10
at three representative stages:
(a, d) the early stage, 
(b, e) the stage of the first star formation, and 
(c, f) the stage of $M_\mathrm{\star, max} = 10M_\sun$.
The color scale shows the column density in the $x$-direction (upper
figures) and the $z$-direction (lower figures).
The blue circles and their associated lines show the positions of the sink particles
and their loci, respectively.
\label{colden_HT10F10.eps}
}
\end{figure*}

Model HT10F10 is the same as the previous model, HT10F0, but 
colliding flow with a Mach number of 10 is added. 
Figure~\ref{colden_HT10F10.eps} shows the evolution of model HT10F10.
The upper and lower panels show the column densities along the $x$-
and $z$-directions, respectively. 
Figures~\ref{colden_HT10F10.eps}(a) and \ref{colden_HT10F10.eps}(d)
show the cloud in the early phase.  The column density along the
$x$-direction (Figure~\ref{colden_HT10F10.eps}(a)) is similar to that of
model HT10F0 (Figure~\ref{colden_HT10F0.eps}(a)), and the gas
compression due to the colliding flow is shown in the column
density along the $z$-direction (Figure~\ref{colden_HT10F10.eps}(d)).
Figures~\ref{colden_HT10F10.eps}(b) and \ref{colden_HT10F10.eps}(e)
show the cloud when the first sink particle forms. 
The colliding flow accumulates filaments to form a sheet at
$x \simeq 0$, and the dense gas of the sheet leads to star formation 
at an earlier time than for the previous model. 

Figures~\ref{colden_HT10F10.eps}(c) and \ref{colden_HT10F10.eps}(f)
show the cloud at the stage of $M_\mathrm{\star, max} = 10\,M_\sun$.
The thickness of the sheet increases because the accumulated filaments
penetrate the sheet.
In the comparison of Figure~\ref{colden_HT10F0.eps}(c) and 
Figure~\ref{colden_HT10F10.eps}(c), 
model HT10F10 exhibits more prominent fine filaments. 
\revise{The filamentary structure is enhanced by the self-gravity,
  while the structure formation is primarily dominated by the
  turbulence.  The colliding flow increases the density in the sheet,
  increasing the effects of the self-gravity.
  We confirmed that the dense filaments with widths of $\sim 0.1$~pc
  have the Jeans length less than $0.1$~pc, indicating that the
  self-gravity also contributes to the structure formation mainly on a scale
  of the filament width.  }
The number of sink particles is roughly the same as that of model
HT10F0 (Table~\ref{table:model-parameters}),
but the elapsed time is shorter than that in model HT10F0 due to the
high accretion rate, which indicates that the colliding flow promotes star formation.

The column density along the $z$-direction shows that 
dense filaments lie parallel to the sheet and 
low-density striations lie perpendicular to the sheet.
The low-density striations are oriented parallel to the accretion flow on
the sheet.
The column density and the velocity are in agreement with recent observations of
\citet{Palmeirim13}.  After finding
striations perpendicular to the filament
in the B211/L1495 region in the Taurus molecular cloud , they suggest that 
gas may be accreting along the striations onto the main filament.

\subsection{Dependence on the colliding flow}
\label{sec:Dependence_of_the_colliding_flow}
\begin{figure*}
\epsscale{1.}
\plotone{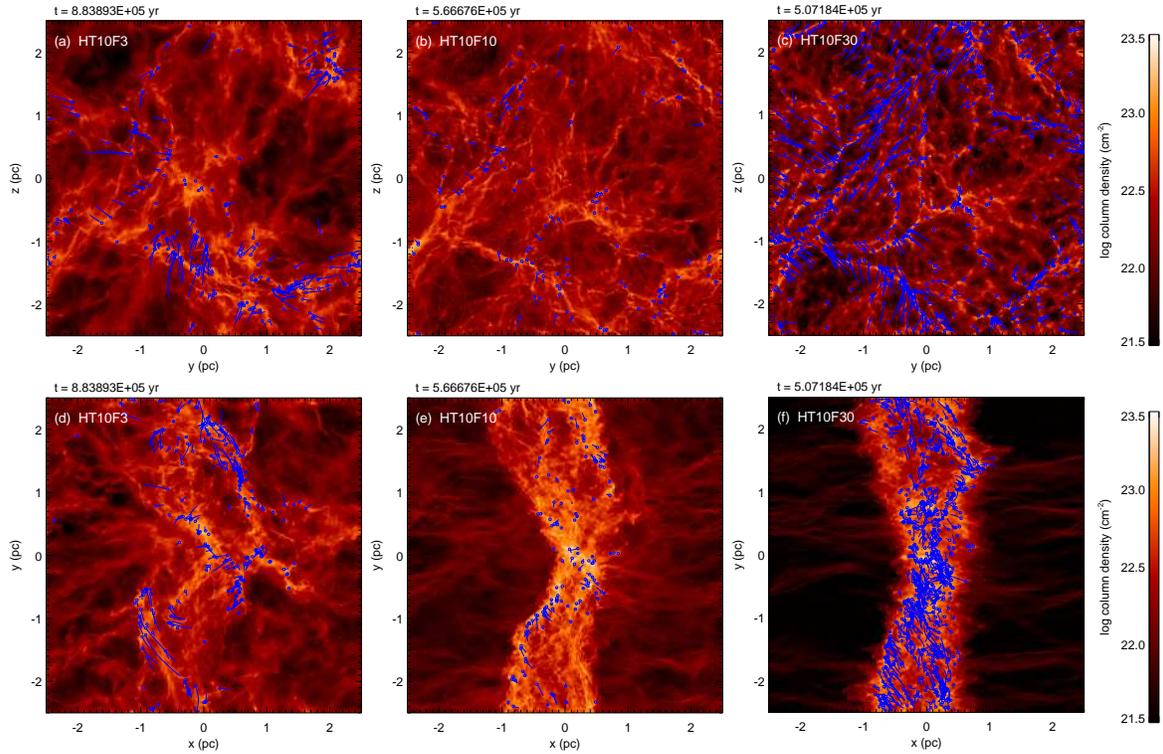}
\figcaption[]{ 
Comparison between the high-density models with a common initial
turbulence of ${\cal M}_t = 10$ and with different strengths of colliding flows
at the stages of $M_\mathrm{\star, max} = 10M_\sun$.
The left, middle, and right figures show models with 
${\cal M}_f = 3, 10,$ and 30, respectively.
The color scale shows the column density in the $x$-direction (upper
figures) and the $z$-direction (lower figures).
The blue circles and their associated lines show the positions of the sink particles
and their loci, respectively.
\label{colden_F_HT10.eps}
}
\end{figure*}

Figure~\ref{colden_F_HT10.eps} shows the high-density models with different strengths of
colliding flows in column densities along the $x$-direction (upper panels) and
the $z$-direction (lower panels). 
The model with a strong flow, HT10F30, exhibits strong compression of
gas into the plane at $x \simeq 0$, as shown in the column densities
along the $z$-direction.
This leads to fine filaments and active formation of sink particles
compared to the previous model, HT10F10, as 
shown in the column densities along the $x$-direction. 
Model HT10F10 produces 142 sink particles, while model HT10F30 produces
915 sink particles.  

The model with weak colliding flow, HT10F3, does not exhibit a clear sheet structure
at $x \simeq 0$, as shown in Figure~\ref{colden_F_HT10.eps}(d).
This is because the amplitude of the colliding flow is considerably
lower than the mean velocity of the initial turbulence 
(${\cal M}_f < {\cal M}_t$).  The low compression due to the
colliding flow leads to a long elapsed time for the maximum mass of the sink particles to reach $10\,M_\sun$.  This long time brings about 
formation of many sink particles compared to the case for model HT10F10.

\begin{figure*}
\epsscale{1.}
\plotone{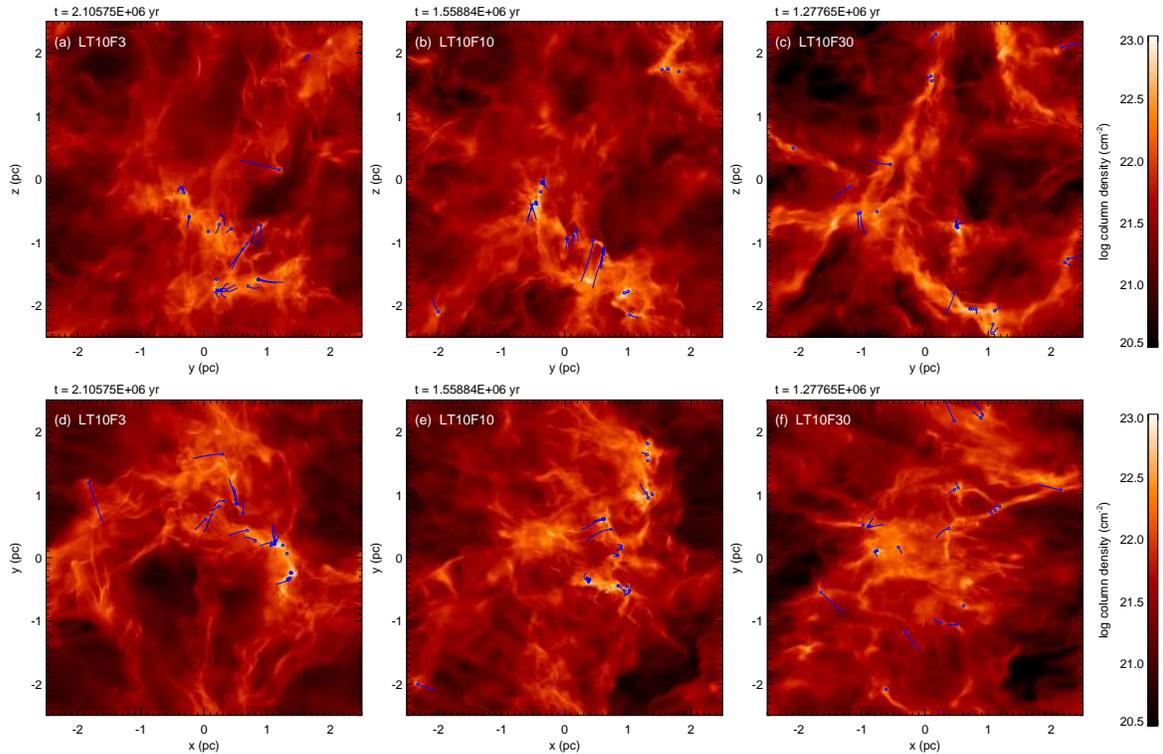}
\figcaption[]{ 
Comparison between the low-density models with a common initial
turbulence of ${\cal M}_t = 10$ and with different strengths of colliding flows
at the stage of $M_\mathrm{\star, max} = 10M_\sun$.
The left, middle, and right figures show models with 
${\cal M}_f = 3, 10,$ and 30, respectively.
The color scale shows the column density in the $x$-direction (upper
figures) and the $z$-direction (lower figures).
The blue circles and their associated lines show the positions of the sink particles
and their loci, respectively.
\label{colden_F_LT10.eps}
}
\end{figure*}

Figure~\ref{colden_F_LT10.eps} compares the low-density models with
different strengths of colliding flows.  In the low-density models,
formation of sink particles begins after the colliding flows pass by
each other.  At the stage of $M_\mathrm{\star, max} = 10\,M_\sun$, the
sheet structures due to colliding flows disappear.

\subsection{Decay of turbulence}

\begin{figure*}
\epsscale{1.0}
\plottwo{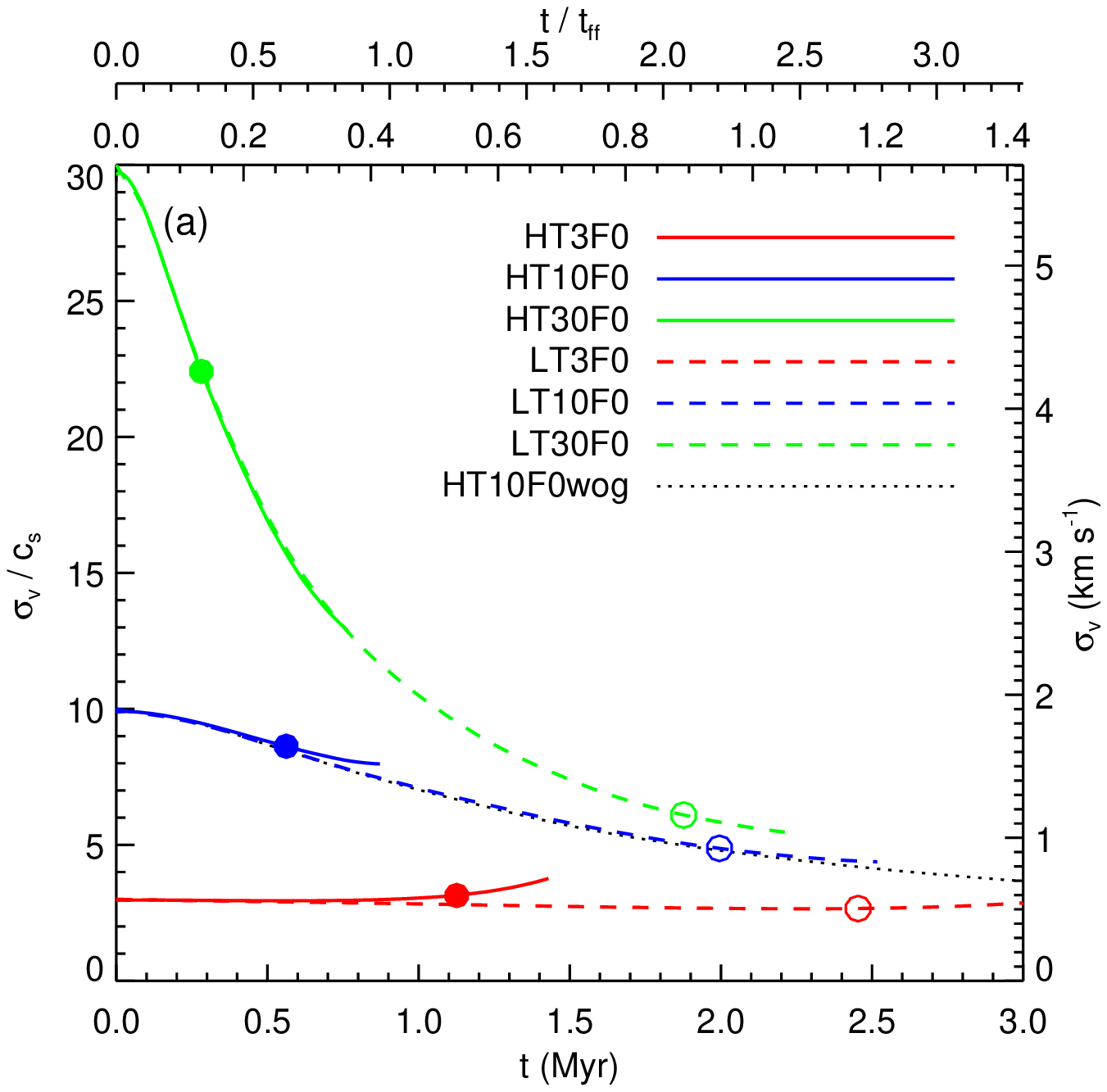}{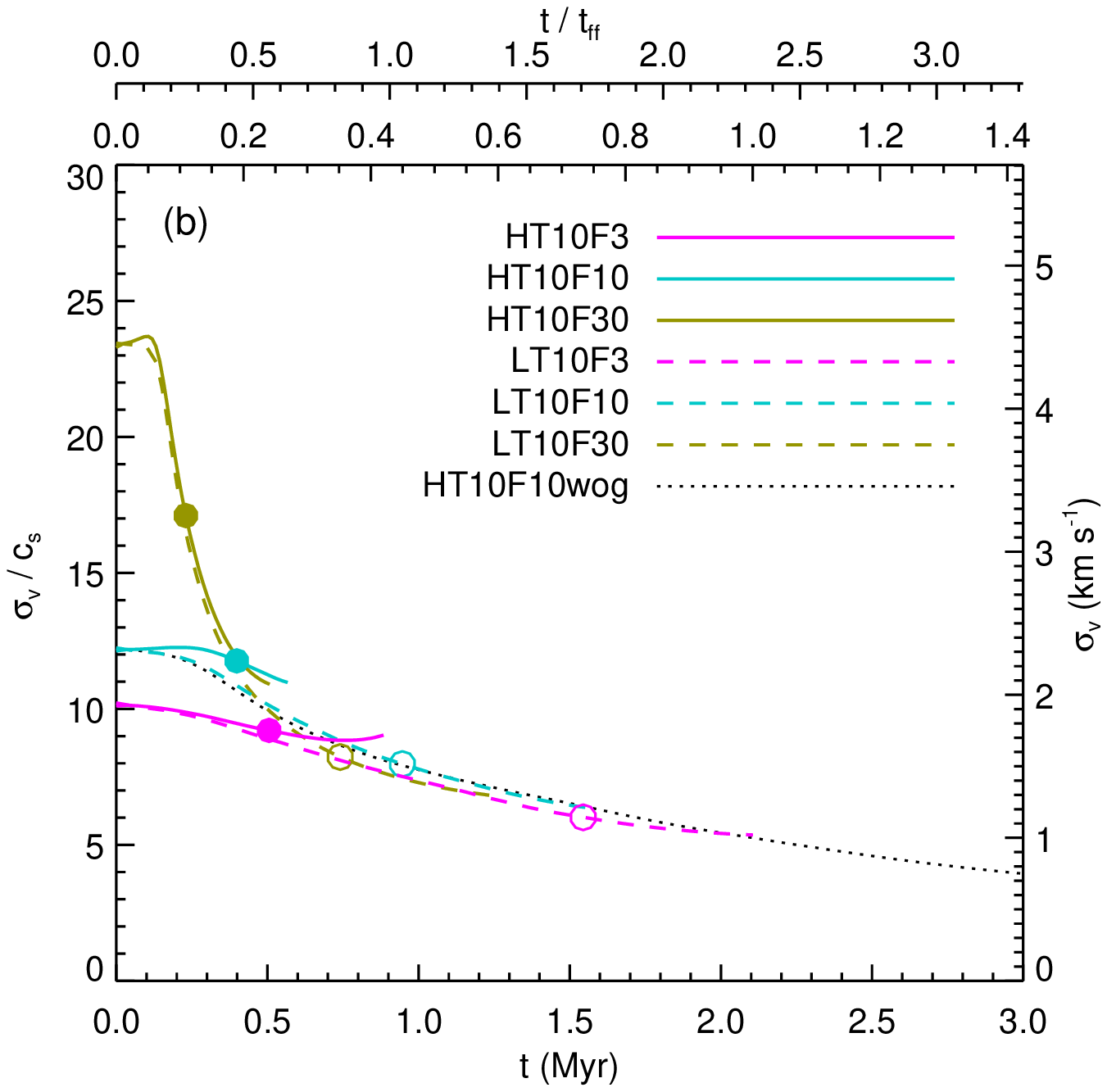}
\figcaption[]{
Velocity dispersions ($\sigma_v$) as functions of time
for all models (a) without and (b) with the
colliding flows.
The solid and dashed curves denote $\sigma_v$ for the high-density and
the low-density models, respectively,
from the initial stages to the stage of $M_\mathrm{\star, max} = 10\,M_\sun$.
The dotted black curves denote the model without self-gravity.
The filled and open circles denote the
first sink particle formation for the high-density and low-density
models, respectively. 
The upper two abscissas show the time normalized by the freefall time of 
the high-density (upper axis) and the low-density models (lower axis).
\label{logugr_plot_hl_lin.eps}
}
\end{figure*}

Figure~\ref{logugr_plot_hl_lin.eps} shows the decay of turbulence 
by measuring the standard deviations of the velocities (velocity dispersion),
$\sigma_v = \langle (\bmath{v}-\langle\bmath{v}\rangle_m)^2 \rangle_m^{1/2}$, 
where $\langle \cdot \rangle_m$ denotes the mass-weighted average over the
computational domain, i.e., $\langle \bmath{v} \rangle_m = \int_V
\rho \bmath{v} \, dV/M_\mathrm{gas,tot}$
and $M_\mathrm{gas,tot} = \int_V \rho dV$. 
The total kinetic energy of gas is therefore evaluated as 
$E_K = M_\mathrm{gas,tot} \sigma_v^2/2$ when $\langle\bmath{v}\rangle_m=0$.
Our numerical scheme conserves the linear momentum of gas within
a truncation error before sink particle formation,
and the mean velocity $\langle\bmath{v}\rangle_m$ therefore has small values compared
to $\sigma_v$. 
Even after sink particle formation, 
the mean velocity remains small, 
$\langle\bmath{v}\rangle_m / \sigma_v \lesssim 10^{-3} - 10^{-2} $ at most.

\begin{figure*}
\epsscale{1.0}
\plottwo{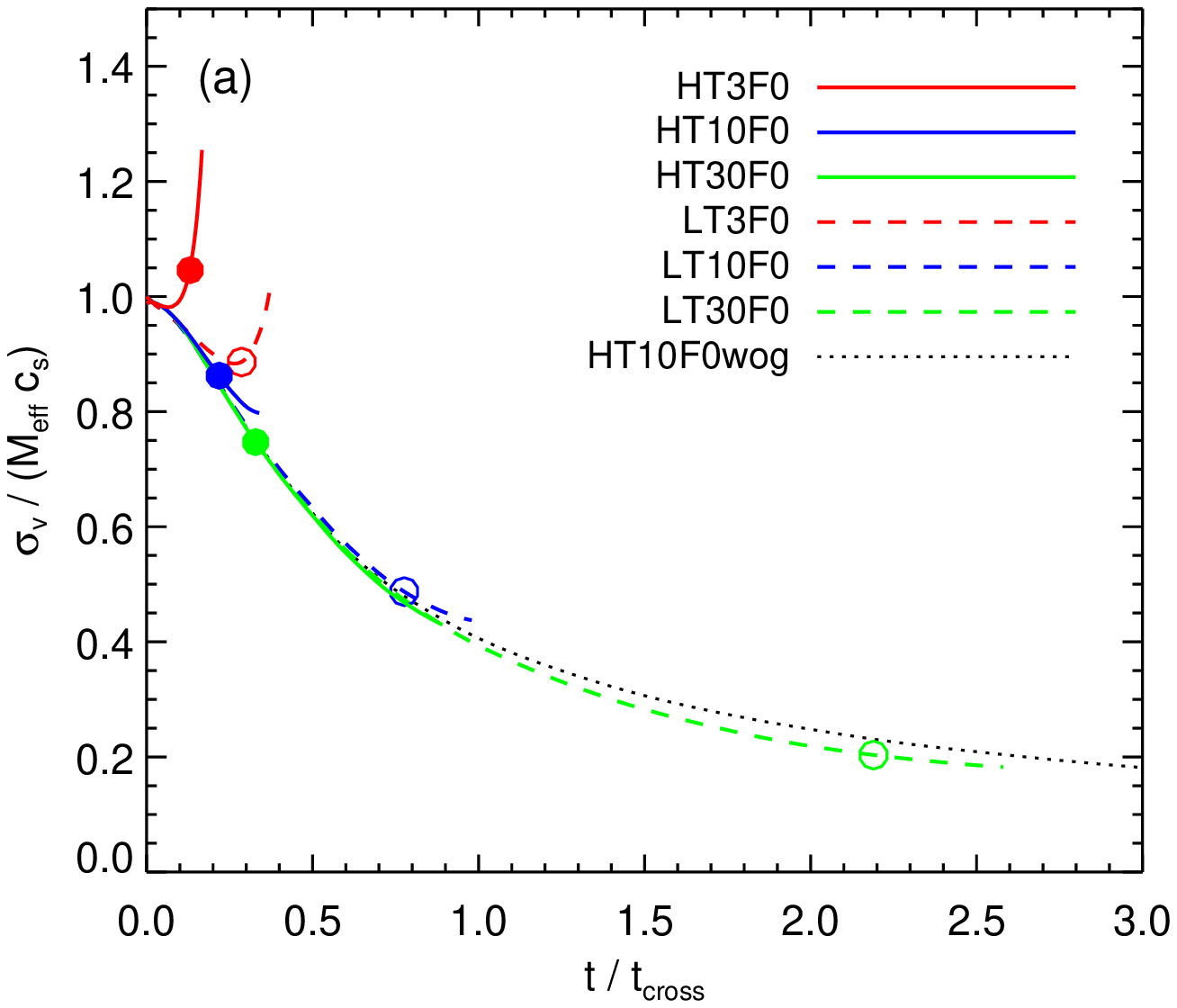}{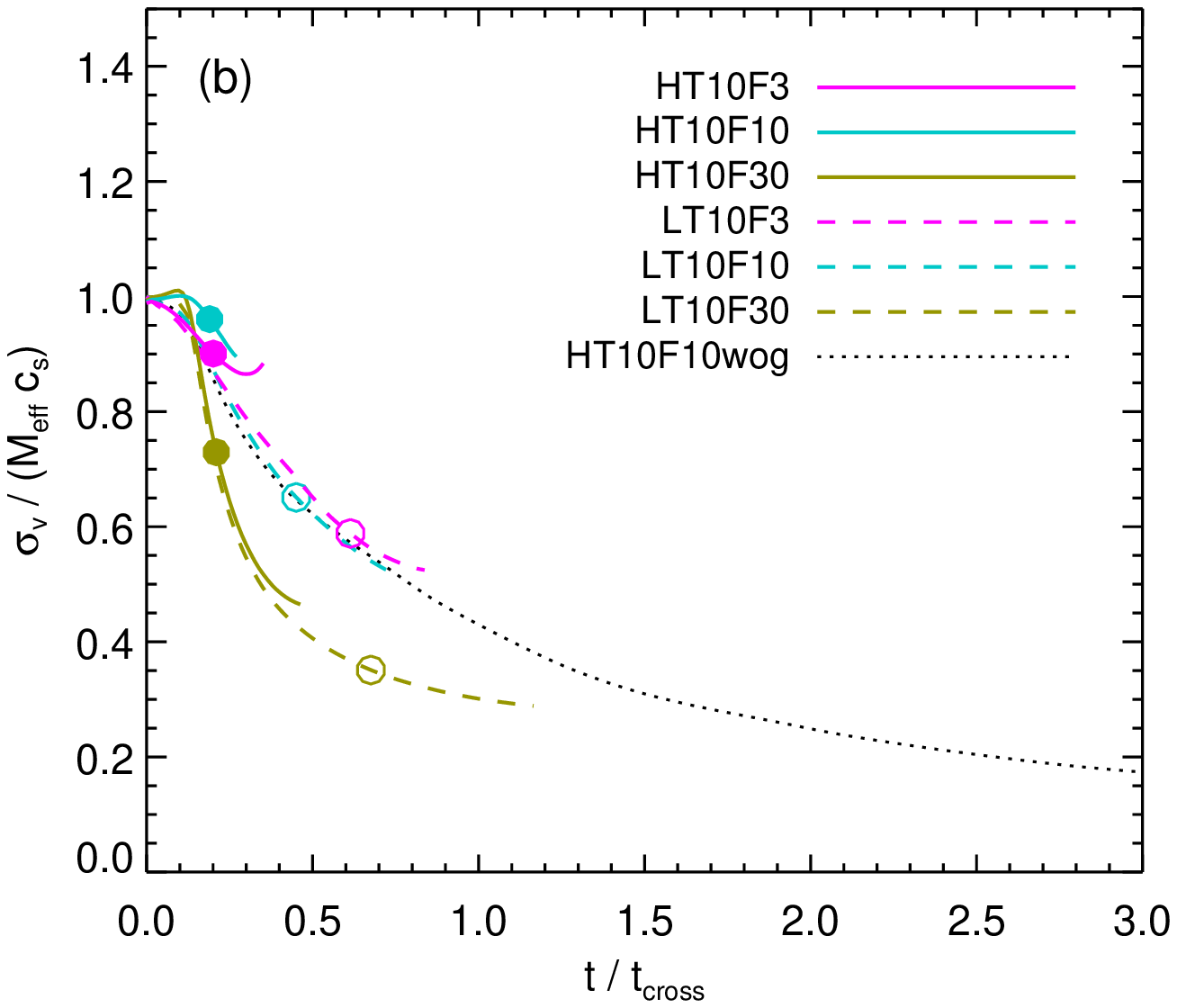}
\figcaption[]{
Same as Figure~\ref{logugr_plot_hl_lin.eps} but 
velocity dispersion and time are normalized by the effective Mach
number ${\cal M}_\mathrm{eff}$ and the crossing time
$t_\mathrm{cross}$, respectively.
\label{logugr_plot_hl_tc.eps}
}
\end{figure*}

For models without colliding flows
(Figure~\ref{logugr_plot_hl_lin.eps}(a)), 
models with a common ${\cal M}_t$ exhibit 
similar decay of turbulence irrespective of their initial densities.
After the stages of formation of the first sink particle, as denoted by
circles, the high-density models tend to have a slightly higher $\sigma_v$ than the
low-density models because of self-gravity. 
The models without colliding flows produce the first sink particle
roughly around $\sim t_\mathrm{ff}$ irrespective of the initial density, 
whereas the time of formation of the first sink particle tends to 
be early when assuming strong turbulence.

When the time and velocity dispersion are normalized by the crossing
time and the effective Mach number, respectively, the evolutions for
the different models converge to that for model MT10F0wog
(non-gravitational model), as shown in
Figure~\ref{logugr_plot_hl_tc.eps}(a).  This indicates that the turbulence
decays in the crossing time scale, $t_\mathrm{cross}$, are in agreement with
\citet{Maclow98}, \citet{Maclow99}, and \citet{Ostriker01}.  The
high-density models and the low-density model LT3F0 undergo gravitational
collapse to form sink particles before the initially assumed turbulence decays considerably.

Figure~\ref{logugr_plot_hl_lin.eps}(b) and
Figure~\ref{logugr_plot_hl_tc.eps}(b) show velocity dispersions as
functions of time for the models with colliding flows.  
The colliding flows contribute to $\sigma_v$, and 
$\sigma_v$ at $t=0$ depends on ${\cal M}_f$ when comparing
models with the same ${\cal M}_t$, as shown in 
Figure~\ref{logugr_plot_hl_lin.eps}(b).
All the models tend to converge to the locus
of the non-gravitational model as time proceeds (Figure~\ref{logugr_plot_hl_lin.eps}(b)).
This tendency indicates that colliding flows decay faster than do isotropic
turbulent flows.  For example, models with ${\cal M}_f = 30$ show
rapid decreases in $\sigma_v$, and the timescales of the decreases are
smaller than $t_\mathrm{cross}$ (Figure~\ref{logugr_plot_hl_tc.eps}(b)),
where $\sigma_v$ decreases by a factor of 50\% in a time of $0.3 t_\mathrm{cross}$.
The rapid decreases in $\sigma_v$ are responsible for the shock
waves formed by the colliding flows.
The colliding flow brings forward the first sink particle formation,
and the elapsed time until the first sink particle formation
is shorter than the freefall time irrespective of the initial densities.

\subsection{Compressibility of velocity fields}
\label{sec:compressibility_of_velocity_fields}

A velocity field can be separated into solenoidal (transverse) and
compressive (longitudinal)
components by applying a Helmholtz decomposition.  
The decomposition of the velocity field has also been used to discuss turbulence
\citep[e.g.,][]{Kitsionas09,Federrath10}.
We performed the velocity decomposition and estimated
$E_\mathrm{trans}$ and $E_\mathrm{long}$
according to 
Equations~(\ref{eq:e_trans}) and (\ref{eq:e_long}), where 
$E_\mathrm{trans}$ and $E_\mathrm{long}$ are the powers of the 
transverse and longitudinal velocity components, respectively
(see Appendix \ref{sec:decomposition}).
Figure~\ref{helmholtz_plot.eps} shows
$E_\mathrm{long}/E_\mathrm{tot} $ as functions of
time, where $E_\mathrm{tot} = E_\mathrm{long}+E_\mathrm{trans}$.

\begin{figure*}
\epsscale{1.0}
\plottwo{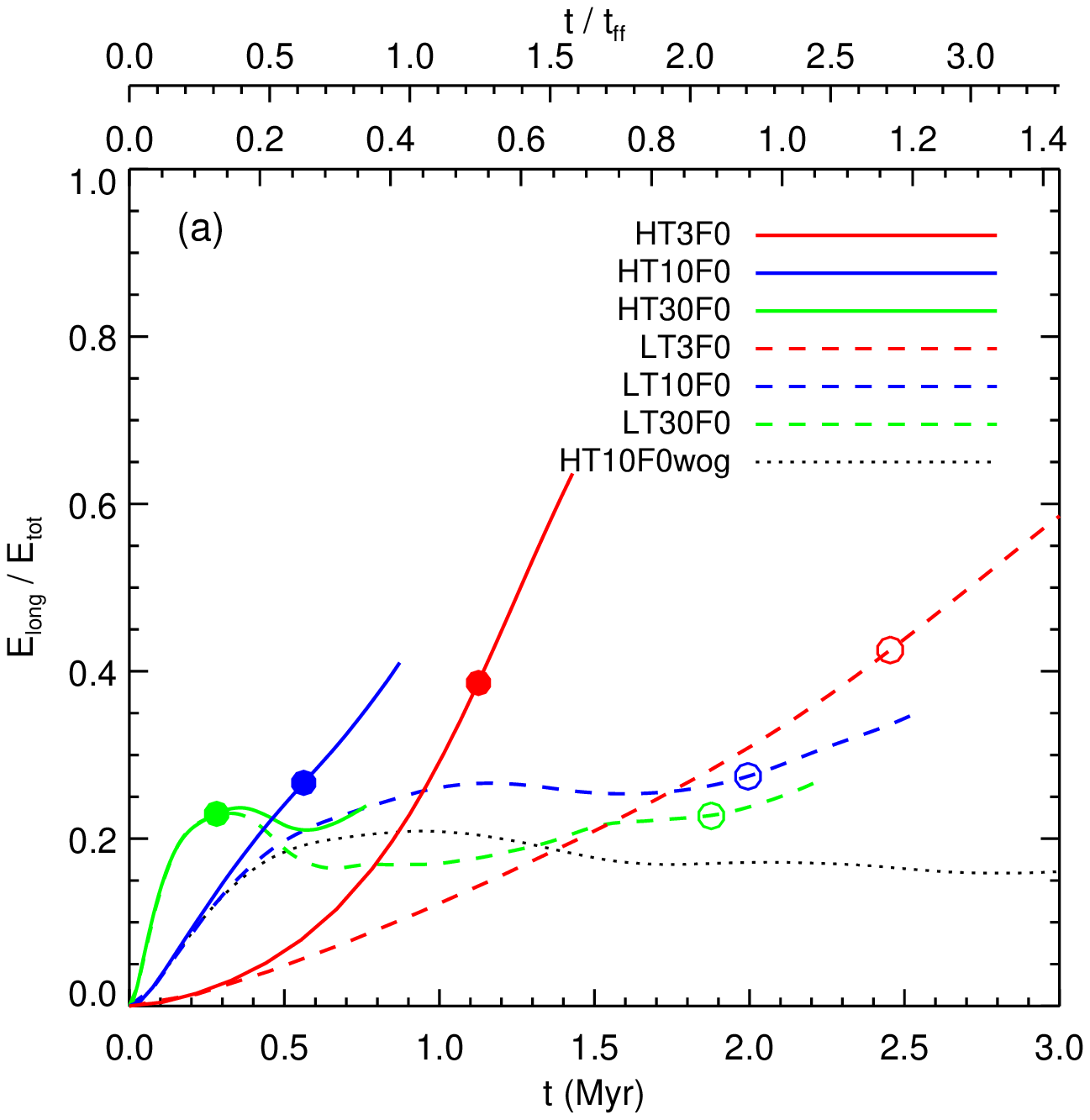}{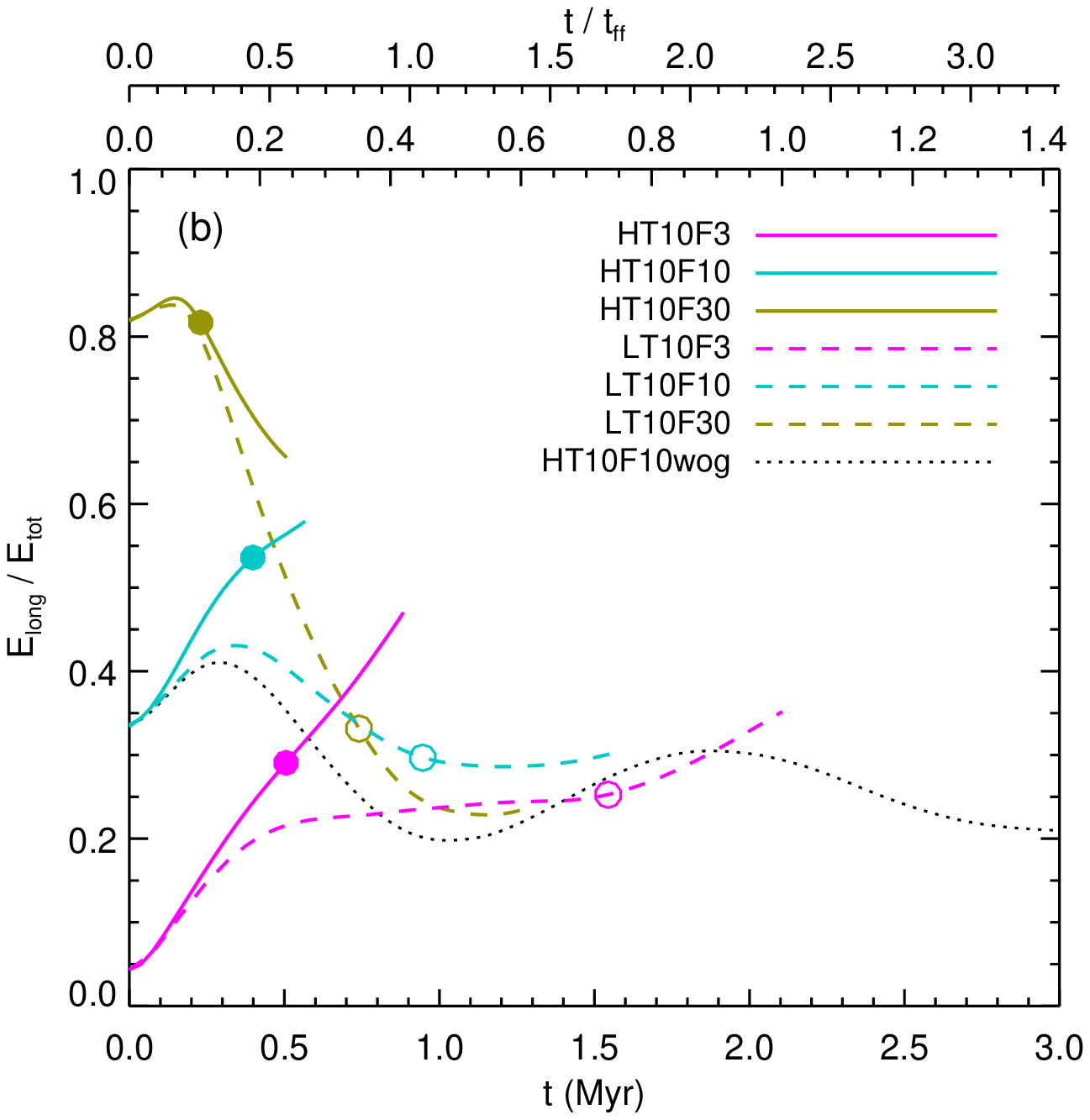}
\figcaption[]{
Ratio of longitudinal power to total power,
$E_\mathrm{long}/E_\mathrm{tot}$,
as functions of time
for the models (a) without and (b) with the colliding flows. 
The solid and dashed curves denote the ratios of the powers for the
high-density and low-density models, respectively.
The dotted black curves denote the model without self-gravity.
The filled and open circles denote the
first sink particle formation for the high-density and low-density
models, respectively. 
The upper two abscissas show the time normalized by the freefall time of 
the high-density (upper axis) and low-density models (lower axis).
\label{helmholtz_plot.eps}
}
\end{figure*}

The models without colliding flow show
$E_\mathrm{long} / E_\mathrm{tot} = 0$ at the initial stages
($t = 0$),
because the initial velocity field is incompressible.
As time proceeds, $E_\mathrm{long} / E_\mathrm{tot}$ increases. This
increase indicates that the compressibility of the velocity field increases.
In the early stages, models with the same turbulent Mach number
and a different initial density exhibit similar
increases in $E_\mathrm{long} / E_\mathrm{tot}$.
Models HT10F0 and LT10F0 show increases in $E_\mathrm{long} / E_\mathrm{tot}$
at a rate similar to the non-gravitational model HT10F0wog.
Models with strong
turbulence tend to show a rapid increase in $E_\mathrm{long} /
E_\mathrm{tot}$, and the increase rates are in proportion to the
crossing time, $t_\mathrm{cross}$.
These increases indicate that evolution in the early stages is controlled
mainly by the turbulence.  

In the later stages, the self-gravitational models HT10F0 and
LT10F0 exhibit higher ratios of power than does the non-gravitational
model HT10F0wog. This indicates that
self-gravity increases the compressibility of the velocity field.
The high-density models tend to exhibit more rapid increases in
$E_\mathrm{long} / E_\mathrm{tot}$ than do the low-density models, and
this result underscores the differences in freefall times between them.

The non-gravitational model HT10F0wog, followed for a long time
of $2.5\times 10^7\,\mathrm{yr}$, exhibits considerable
oscillation of $E_\mathrm{long} / E_\mathrm{tot}$ in the range of $0.15
- 0.2$.  
After the velocity becomes subsonic $\sigma_v \lesssim c_s$, 
the oscillation dumps and $E_\mathrm{long} / E_\mathrm{tot} \simeq
0.15$ is shown.

Models with colliding flows exhibit
$E_\mathrm{long} / E_\mathrm{tot} > 0$ at the initial stages
because the colliding flows are compressible and contribute to
$E_\mathrm{long}$.
The high-density models tend to have a high ratio of 
$E_\mathrm{long} / E_\mathrm{tot}$ than do the low-density models.
Moreover, in comparison with the non-gravitational models, the
self-gravitational modes have a high ratio.
These comparisons also indicate that self-gravity enhances 
the compressibility of the velocity field.  

The non-gravitational model HT10F10wog exhibits an oscillation of 
$E_\mathrm{long} / E_\mathrm{tot}$ with decaying amplitude
in the range of $0.2-0.4$ when
$\sigma_v \gtrsim c_s$ (the early part of the epoch is shown in Figure~\ref{helmholtz_plot.eps}),
and it decreases to $\simeq 0.15$ when $\sigma_v \lesssim c_s$.
The amplitude of the oscillation is larger than that for the
non-colliding model HT10F0wog. 
The oscillation is responsible for the oscillation on $E_\mathrm{long}$, while
$E_\mathrm{trans}$ decreases monotonically.

\subsection{Probability distribution functions}
\label{sec:PDF}
It is known that 
the probability distribution function (PDF) of density in isothermal
turbulent gas without self-gravity is well approximated by a
lognormal distribution \citep[e.g.,][]{Vazquez-Semadeni94}.
The lognormal distribution is expressed as 
\citep{Ostriker01,Padoan02}
\begin{equation}
p(s) ds =
\frac{1}{ (2\pi\sigma_s^2)^{1/2} }
\exp\left[
-\frac{1}{2}
\left(
\frac{s-\bar{s}}{\sigma_s}
\right)^2
\right] ds,
\end{equation}
\begin{equation}
s = \ln \left(\frac{\rho}{\rho_0}\right),
\end{equation}
\begin{equation}
\bar{s} = -\frac{\sigma_s^2}{2},
\end{equation}
where 
$\bar{s}$ and $\sigma_s$ denote the average and the standard deviation
of the logarithmic density, $s$, respectively. 
The standard deviation, $\sigma_s$, is found to be a function of the Mach
number for the gas, ${\cal M}$, as
\citep{Padoan97} 
\begin{equation}
\sigma_s^2 = \ln
\left(
1+ \gamma ^2{\cal M}^2
\right),
\label{eq:lognormal-sigma}
\end{equation}
where $\gamma \simeq 0.5$ and ${\cal M} = \langle \left(v- \langle v\rangle\right)^2 \rangle^{1/2}/c_s$, and
$\langle \cdot \rangle$ denotes a volume-weighted average.

\begin{figure*}
\epsscale{1.}
\plotone{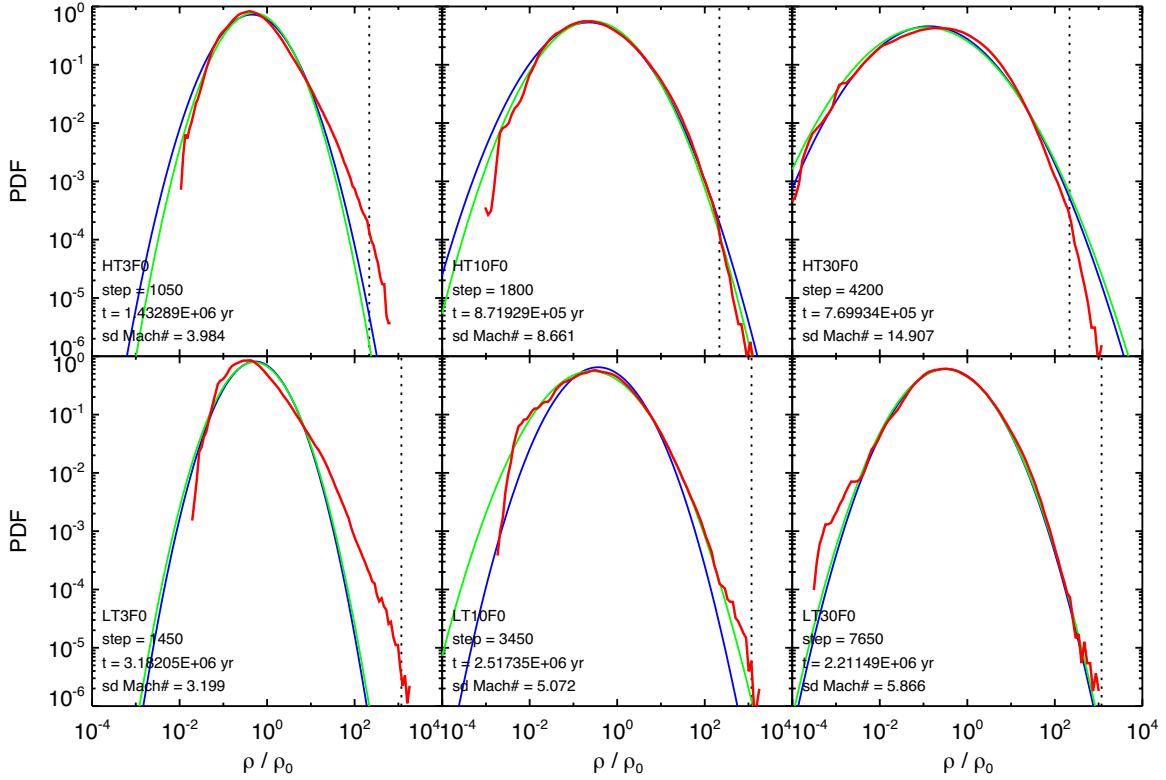}
\figcaption[]{ 
Volume-weighted probability distribution functions (PDFs) for models without 
colliding flow.  The upper and lower figures show PDFs for 
the high-density and low-density models, respectively. 
The left, middle, and right figures show PDFs for models with 
${\cal M}_t = 3, 10, $ and 30, respectively.
The red curves denote the PDFs derived from the density distributions at
the stages of $M_\mathrm{\star, max} = 10\,M_\sun$.
The green and blue curves show the lognormal functions with
the standard deviations derived from 
the density distributions and 
Equation~(\ref{eq:lognormal-sigma}), respectively.
The vertical dotted lines show the threshold densities for the sink
particle, $\rho_\mathrm{sink}$.
\label{pdf_plot_paper_M.eps}
}
\end{figure*}

At the initial stage, the PDFs are expressed by a delta function
because the initial density is constant. As time proceeds, 
the profiles of the PDFs are broadened, and they are well expressed by the
lognormal distributions for all models without colliding
flow in the early phase, typically before sink particles form.
Figure~\ref{pdf_plot_paper_M.eps} shows the volume-weighted PDFs for models without
colliding flow at the stages of $M_\mathrm{\star, max} =10\,M_\sun$.  
For models with ${\cal M}_t = 3$, the PDFs have power-law tails in the high
densities and so deviate from the lognormal functions. 
This indicates that
the cloud structure in high density is affected by self-gravity.
A deviation of PDFs from the lognormal distribution was also
reported by \citet{Klessen00}, \citet{Federrath08}, and \citet{Federrath13} for self-gravitational
isothermal turbulence.
During the evolution, the profiles of the low-density tails
oscillate around the lognormal distributions.  The PDFs
of the snapshots
shown in Figure~\ref{pdf_plot_paper_M.eps} therefore deviate from the
lognormal distribution at low densities. 

In Figure~\ref{pdf_plot_paper_M.eps},
two fitting curves are plotted for each model based on the lognormal functions.
The standard deviation of the logarithmic density is derived from the
density distribution (green curve), and 
the standard deviation is derived by using
Equation~(\ref{eq:lognormal-sigma}) (blue curve).
In the case of non-gravitational driven turbulence, 
\citet{Konstandin12} reported $\gamma = 1/3$ for solenoidal forcing
and $\gamma =1$ for compressive forcing \citep[see][]{Price11}.
We adopt $\gamma=1/2$ here because it provides a better fit to the data than do the
other values.
These two lognormal fitting curves roughly coincide and indicate that
Equation~(\ref{eq:lognormal-sigma}) is satisfied well in the stages examined here, even in the
case of self-gravitating gas.

\begin{figure*}
\epsscale{1.}
\plotone{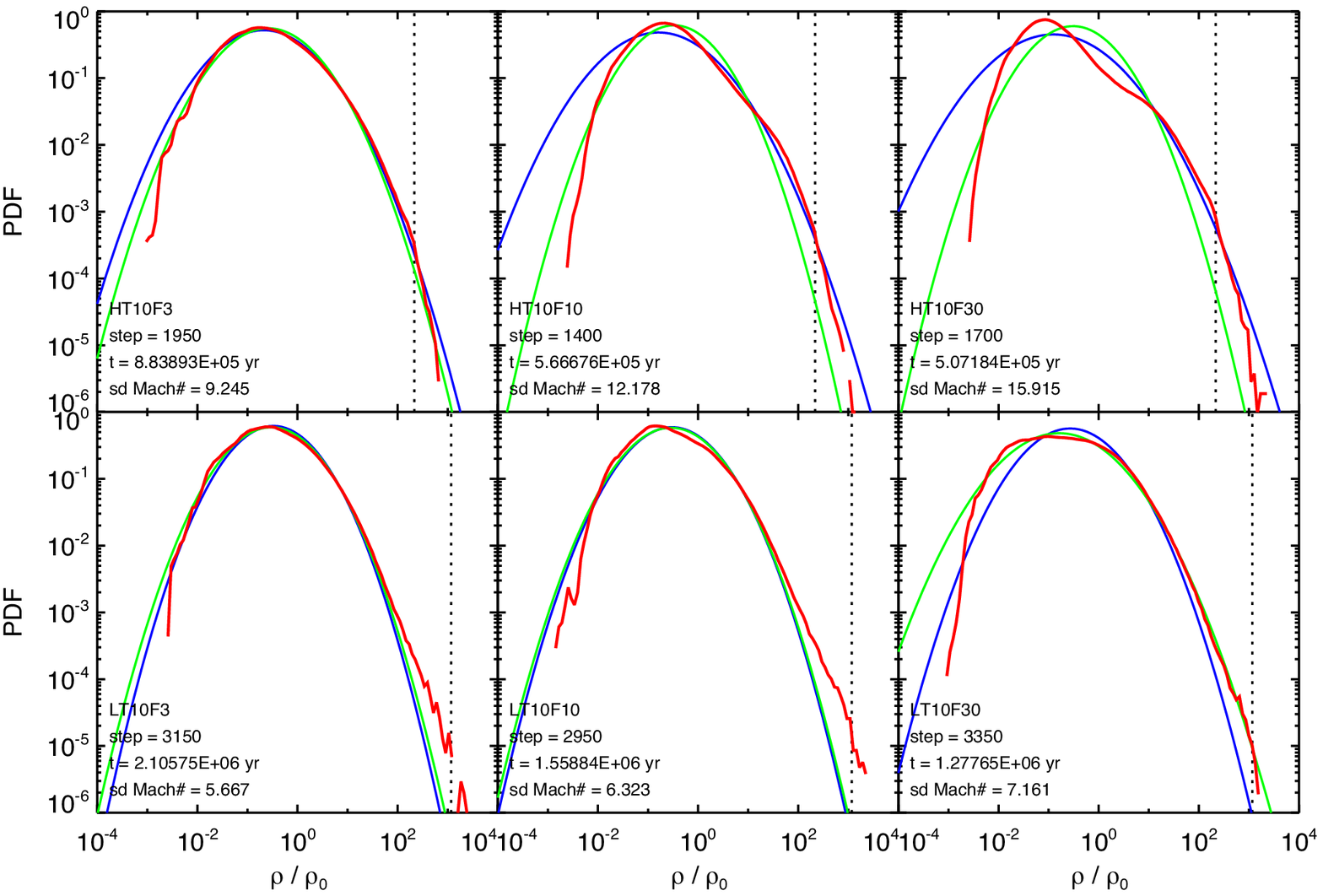}
\figcaption[]{ 
Volume-weighted probability distribution functions (PDFs) for models with a common
initial turbulent strength ${\cal M}_t = 10$ and 
different strengths of colliding flow ${\cal M}_t$.  
The upper and lower figures show PDFs for 
the high-density and low-density models, respectively. 
The left, middle, and right figures show PDFs for models with 
${\cal M}_f = 3, 10, $ and 30, respectively.
The red curves denote the PDFs derived from the density distributions at
the stages of $M_\mathrm{\star, max} = 10\,M_\sun$.
The green and blue curves show the lognormal functions with
the standard deviations derived from 
the density distributions and 
Equation~(\ref{eq:lognormal-sigma}), respectively.
The vertical dotted lines show the threshold density for the sink
particle, $\rho_\mathrm{sink}$.
\label{pdf_plot_paper_F.eps}
}
\end{figure*}

Figure~\ref{pdf_plot_paper_F.eps} shows the volume-weighted PDFs for models with
colliding flow.  In the cases with strong colliding flow, 
especially in the high-density models, 
the PDFs deviate considerably from the lognormal distributions around
the peaks. 
This deviation is attributed to the sheet structures of the clouds caused by the 
colliding flow, 
as shown in 
Figures~\ref{colden_F_HT10.eps}(e) and \ref{colden_F_HT10.eps}(f).
Moreover, the two fitting curves, which show significantly different distributions,
indicate that Equation~\ref{eq:lognormal-sigma}
is no longer satisfied in the cases with strong colliding flow.

PDFs for the column density are shown because
they are often examined in observational studies, \citep[e.g.,][]{Lombardi11}.
\citet{Ostriker01} showed that the PDF for the column density can also be fitted by lognormal functions.

\begin{figure*}
\epsscale{1.0}
\plottwo{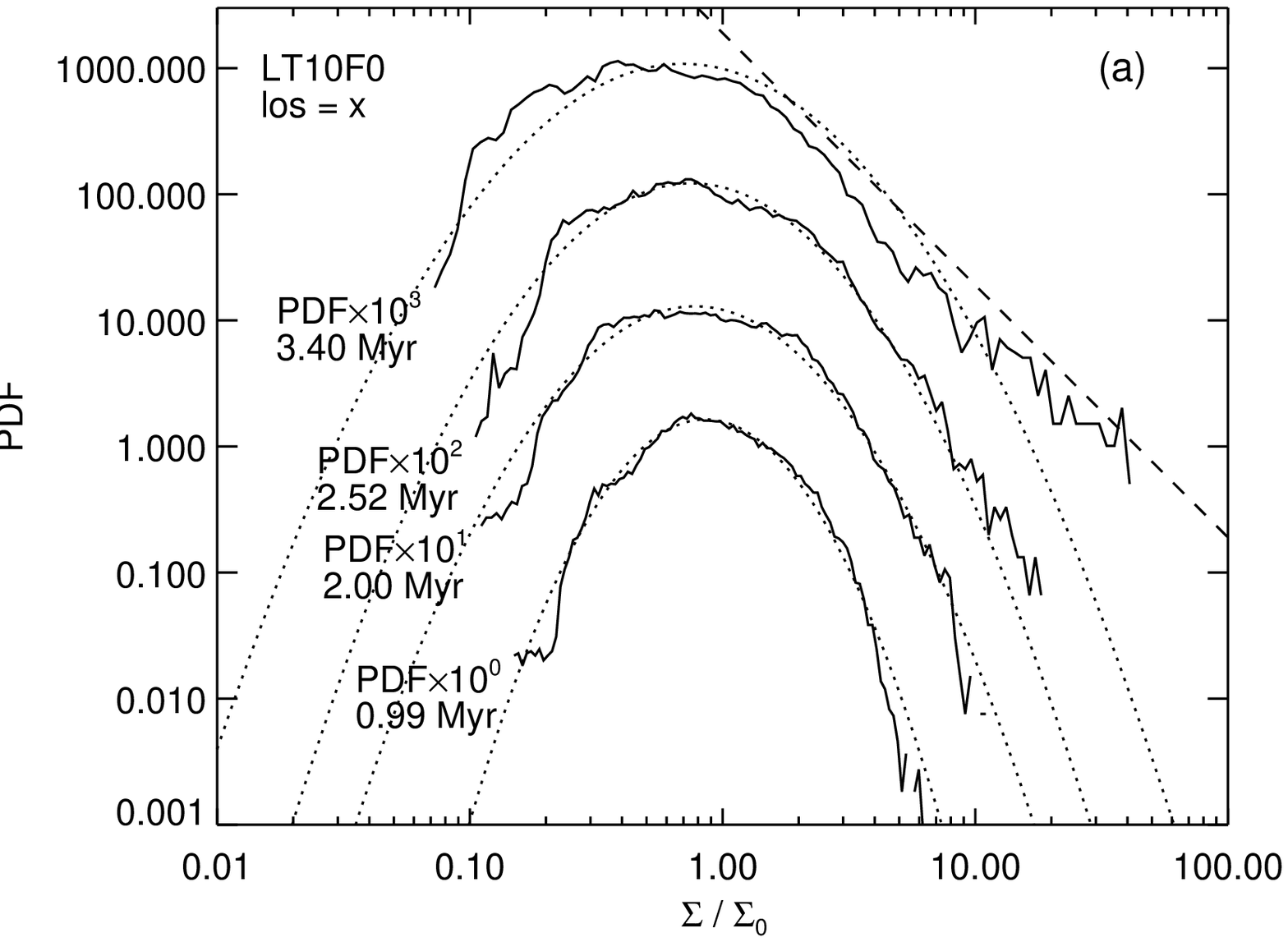}{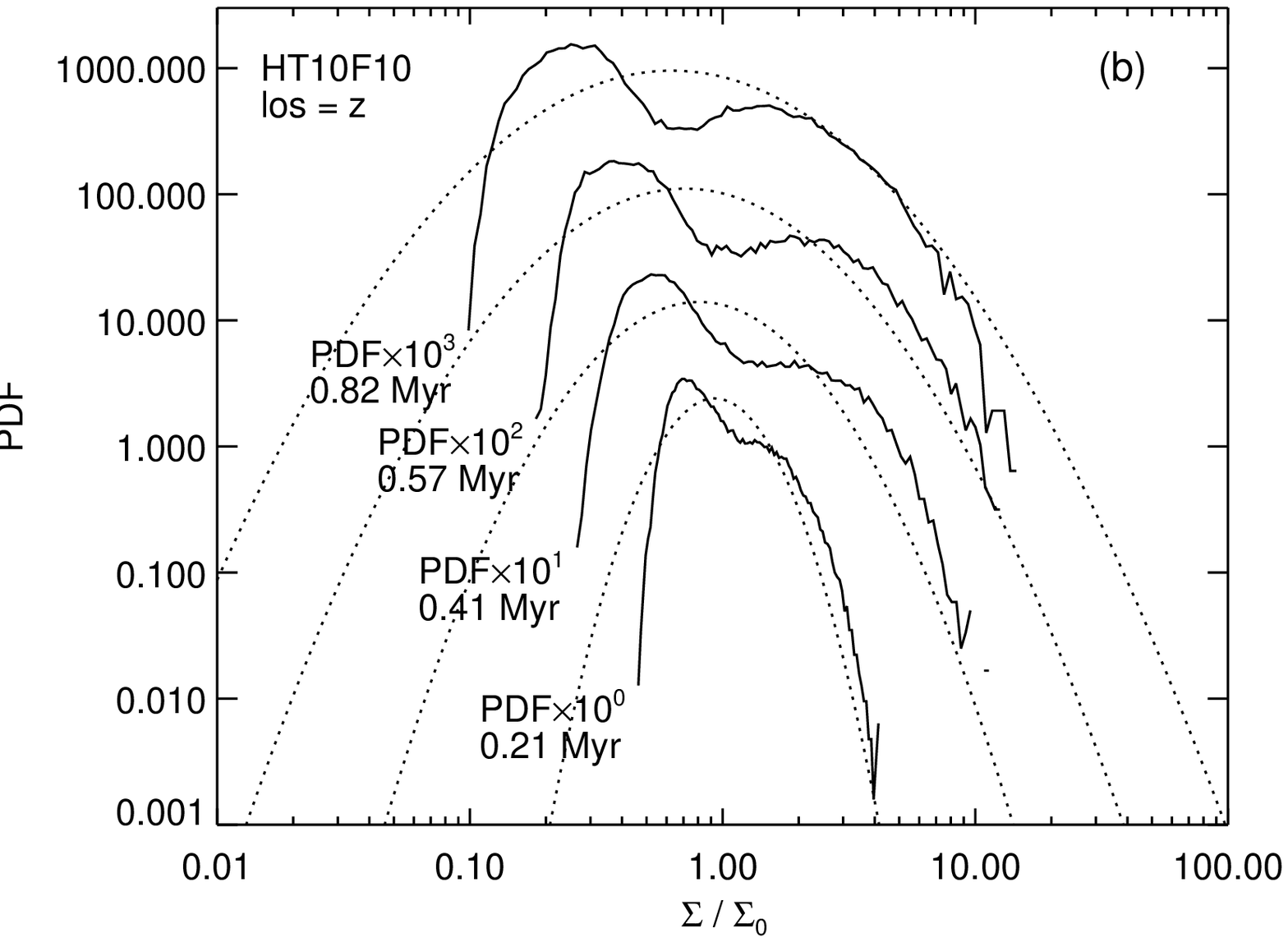}
\figcaption[]{ 
Evolution of PDFs of the column densities for (a) model LT10F0
and (b) model HT10F10.
The column densities are estimated along (a) the $x$-direction
and (b) the $z$-direction.
The solid lines show the PDFs at the four stages, from bottom to top:
the relatively early stage, the stage of the first sink
particle formation, the stage of $M_\mathrm{\star, max} = 10\,M_\sun$, 
and the stage of $M_\mathrm{\star, max} = 50\,M_\sun$.
The times for the stages are shown at the left of the PDF curves.
In order to plot all of the PDFs, the plots are offset
from each other by a factor of 10 in the vertical direction. 
The column densities on the abscissa are normalized by the initial values.
The dotted curves are the fitting curves of lognormal functions, which are
determined by the mean and the standard deviation of the column
densities.
In panel (a), the broken line shows the relationship of $\mathrm{PDF}\propto \Sigma^{-2}$.
\label{pdf_plot_nsteps.eps}
}
\end{figure*}

Figure~\ref{pdf_plot_nsteps.eps}(a) shows the 
PDFs for the column density for model LT10F0. 
The density PDF for Model LT10F0 has a little excess from the
lognormal function in the high densities 
(the bottom panel in the middle of Figure~\ref{pdf_plot_paper_M.eps}).
The profiles of the column density
PDFs are bumpy compared to the density PDFs because the sampling numbers
for the column density are less than those for the density.
For the first three stages shown in Figure~\ref{pdf_plot_nsteps.eps}(a),
each PDF can be fitted well by a lognormal function (dotted curves).
To examine the long-time evolution of the PDF for this model, we follow the
cloud evolution until the stage of $M_\mathrm{\star, max} = 50\,M_\sun$
($t = 3.4\,\mathrm{Myr}$).
The PDF at this stage has a wing at high column density, which can be fitted
by the power law of $\mathrm{PDF} \propto \Sigma^{-2}$. 
The estimated power indexes depend on the models; the power index is
approximated as $-3$ for models HT3F0 and HT10F0, and 
seems to change from $-2$ to $-3$ as the column
density increases for model LT3F0. 
Such a power-law wing in PDFs has often been observed in several
star-forming regions \citep[i.e.,][]{Kainulainen09}, 
as will be discussed in Section~\ref{sec:discussion:PDF}.

Figure~\ref{pdf_plot_nsteps.eps}(b) shows the column density PDFs for the colliding model
HT10F10.  The line of sight is the $z$-direction, which is perpendicular
to the colliding flow.  Along this line of sight, the sheet cloud appears
in the edge-on view.  The PDFs highly deviate from lognormal
functions and exhibit two peaks, which are responsible for 
the sheet cloud with a high column density and the colliding flow with
a low column density.
We confirmed that, when the line of sight is taken to be the face-on direction,
the PDFs can be fitted by a lognormal function.
In summary, the features of the PDFs for the column density depend on the line of sight, and they can highly deviate from a lognormal function
when the cloud is disturbed by a large-scale flow.

\subsection{Mass distribution for sink particles}

\begin{figure*}
\epsscale{1.0}
\begin{center}
\plottwo{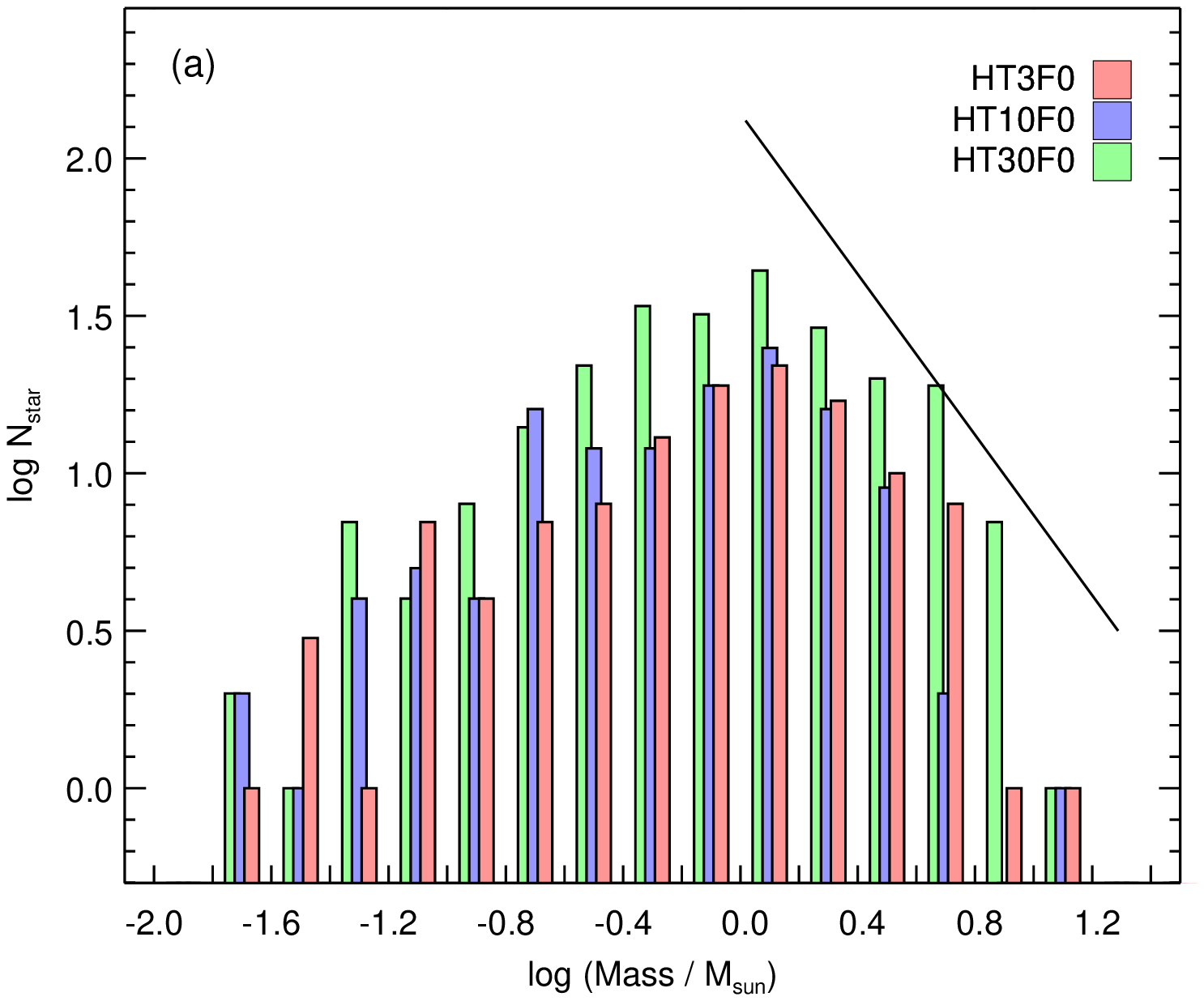}{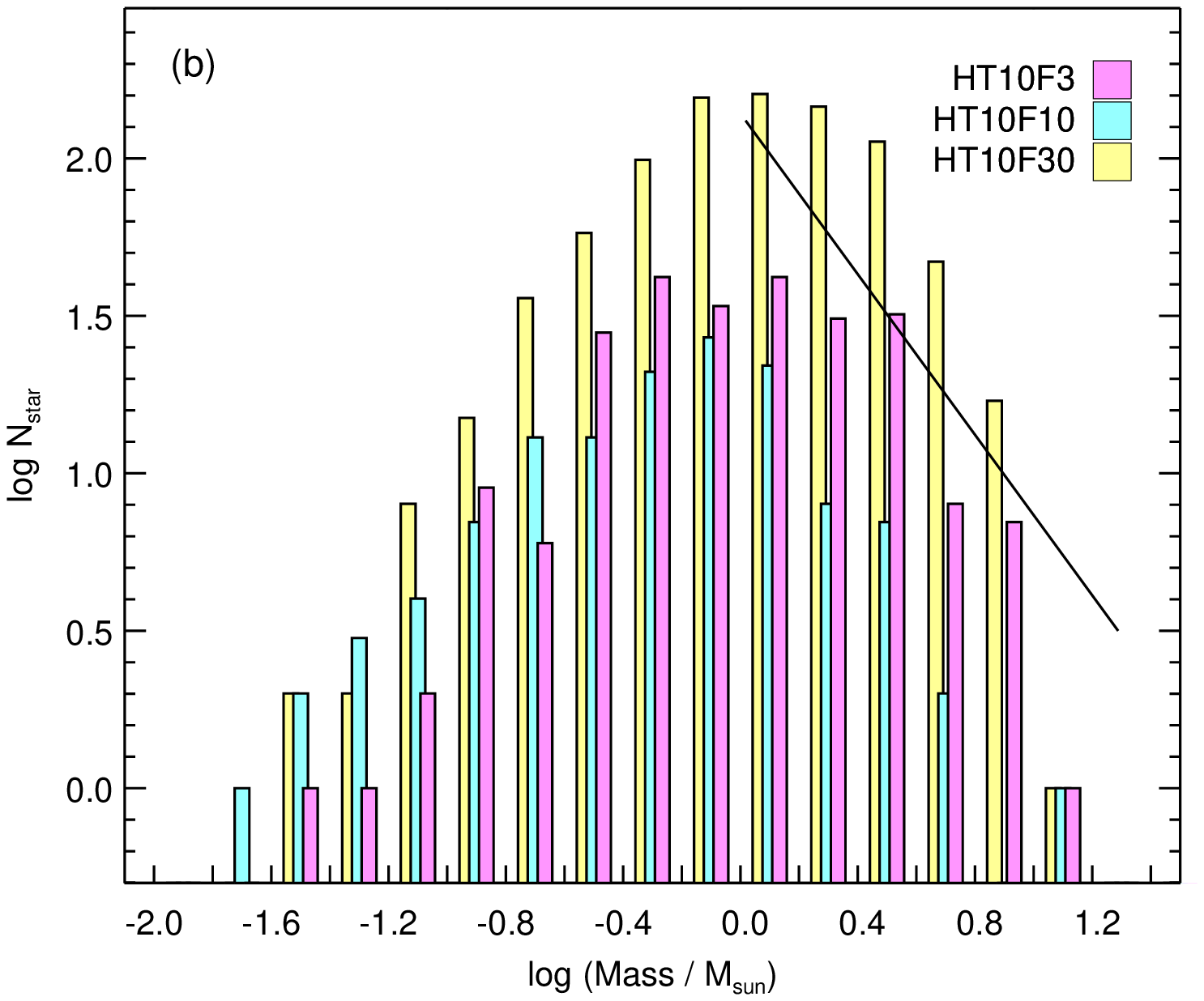}
\plottwo{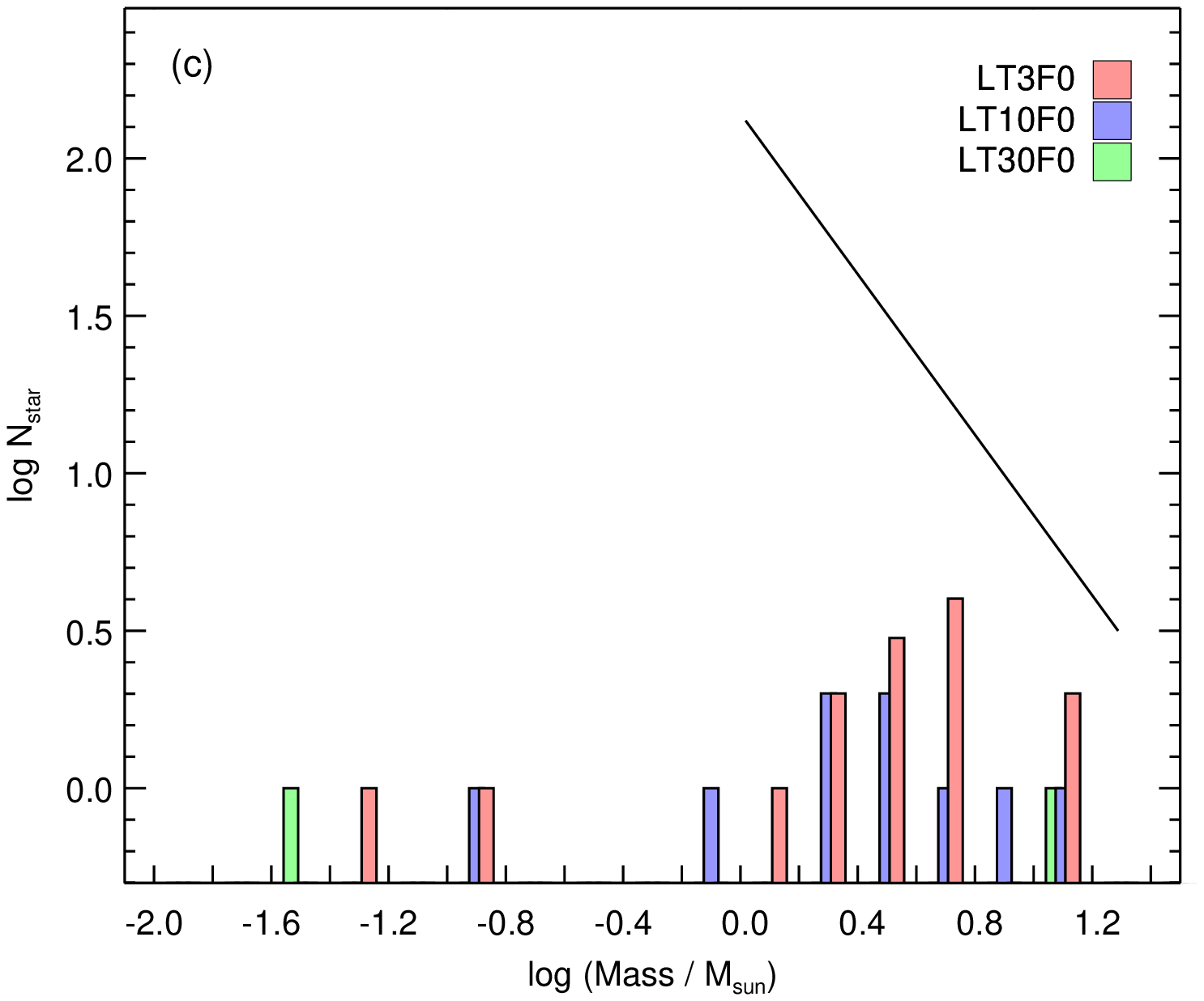}{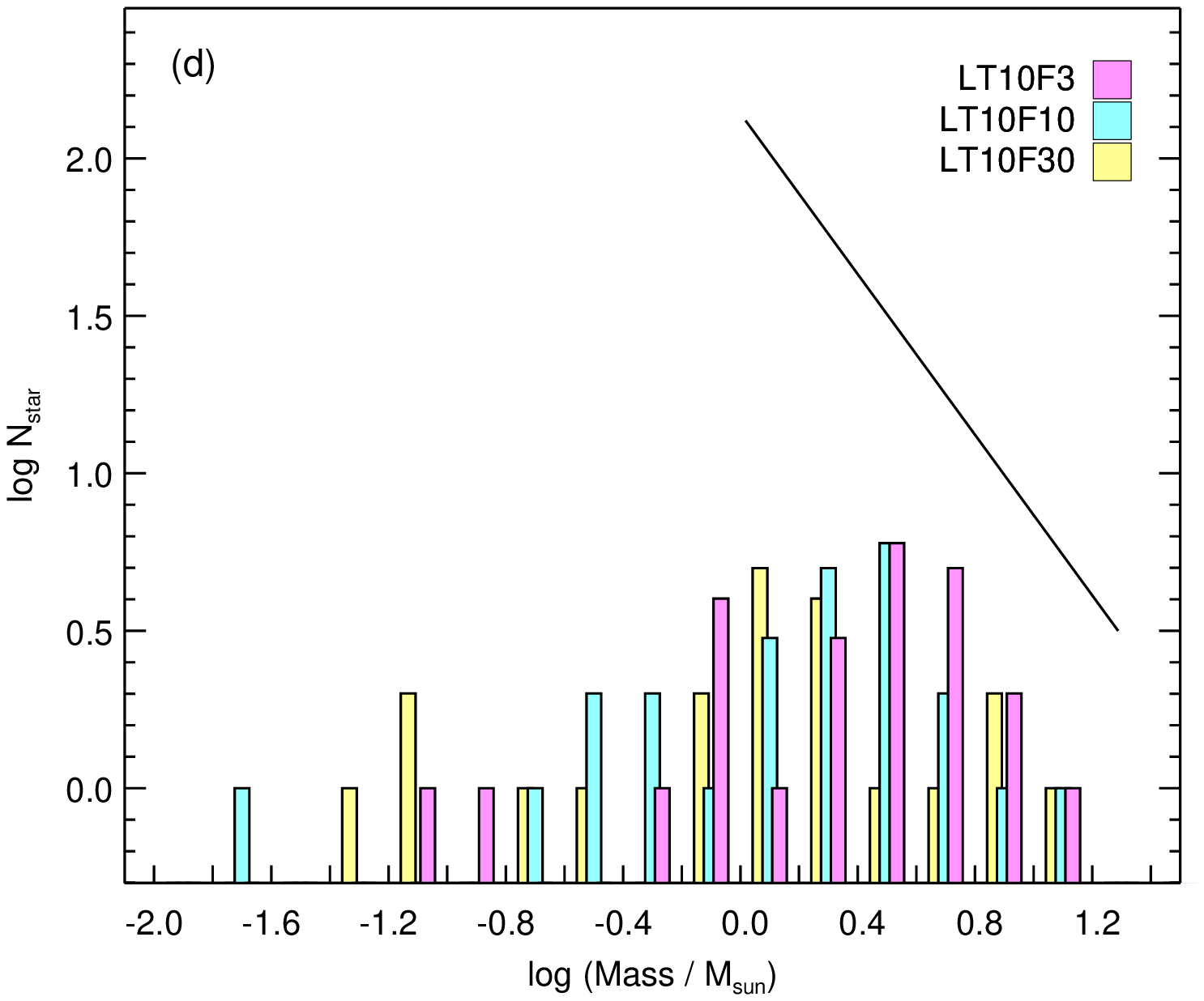}
\end{center}
\figcaption[]{ 
Histograms of the masses of sink particles for 
(a, b) the high-density models and 
(c, d) the low-density models
at the stages of $M_\mathrm{\star, max} = 10\,M_\sun$.
The relationship of 
$N_\mathrm{star}(M_\mathrm{star}) \propto M_\mathrm{star}^{-1.35}$
\citep{Salpeter55}
is plotted by the solid line in
each panel for comparison.
The bin size of the mass is set at $\log 0.2$.
\label{pmass-histogram.eps}
}
\end{figure*}

Figure~\ref{pmass-histogram.eps} shows the mass distribution for 
sink particles in the histograms for all models.  
The high-density models exhibit clear features in the histograms,
whereas the low-density models do not because of the small number of 
sink particles produced.  
For the high-density models, the mass distribution has a peak around
$\sim 1 M_\sun$ and has tails in both the high masses and low
masses. 
For all the high-density models except model HT10F30, 
the tails can be roughly fitted by the relationship 
$N_\mathrm{star}(M_\mathrm{star}) \propto M_\mathrm{star}^{-1.35}$,
which is a classical initial mass function (IMF) of 
\citet{Salpeter55}.
Similar tails at high masses were reported 
for a cloud with a different mass of $500M_\sun$ by \citet{Bate09} and
\citet{Bate12}.
Model HT10F30 shows considerably steeper tails than does the fitting line 
at high mass.
This is because this model produces $3.6-6.5$ times more sink
particles than do the other models (models HT10F3 and HT10F10) and 
the masses of the peaks and the maximum masses are roughly
common among the models.

\begin{figure}
\epsscale{1.}
\plotone{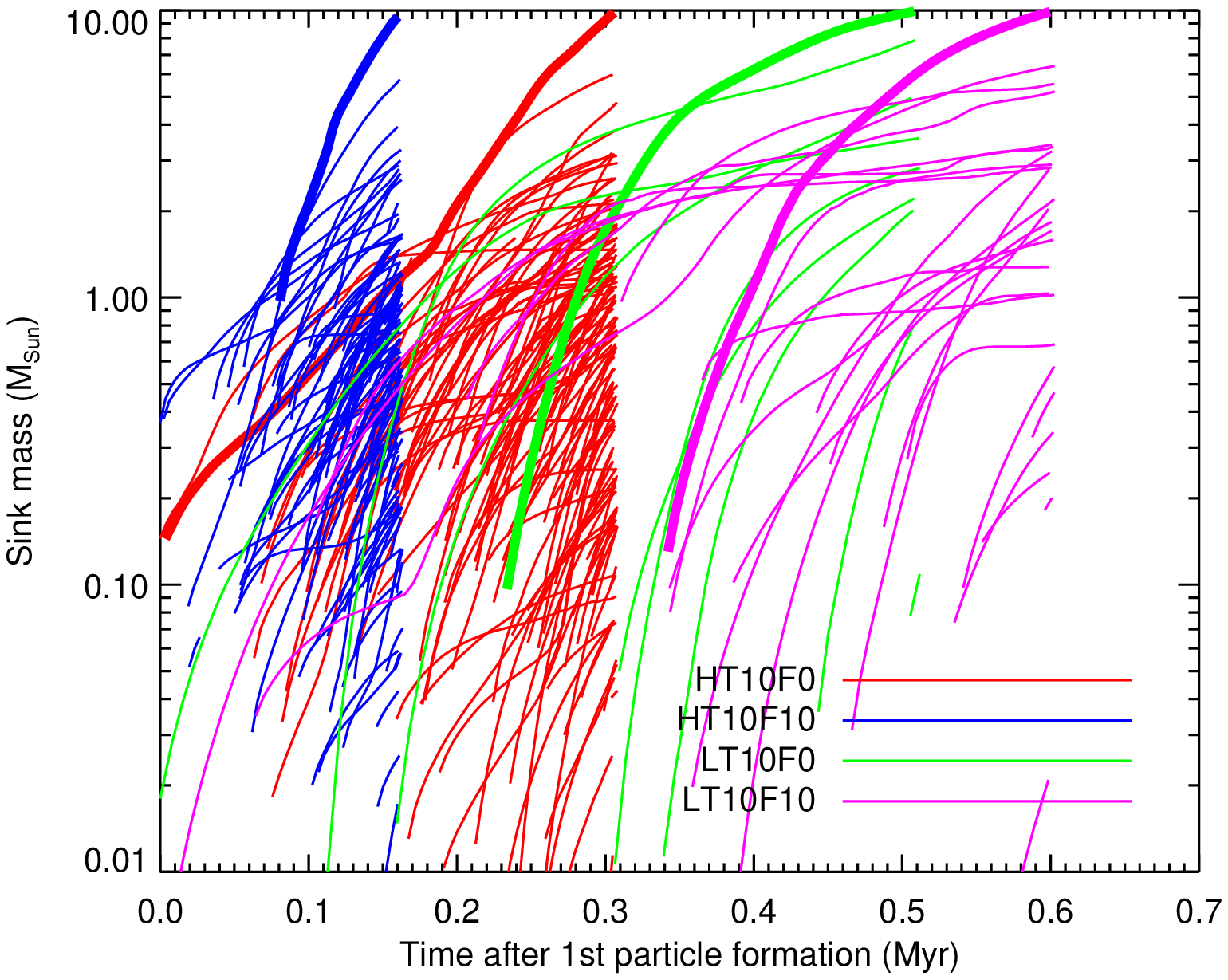}
\figcaption[]{ 
Masses of sink particles as functions of time after the first sink
particle formation for models HT10F0, HT10F10, LT10F0, and LT10F10.
The thick lines show particles reaching $10\,M_\sun$.
\label{time-pmass.eps}
}
\end{figure}

The histograms presented here have a turnover near $\sim 1 M_\sun$,
whereas the IMFs derived from observations in several star-forming regions
tend to have a turnover near 
$\sim 0.07 - 0.5 M_\sun$ \citep[e.g.,][]{Kroupa13}.
A possible interpretation for the discrepancy is that the our
simulations do not spatially resolve the formation of very low mass stars
whose masses are $\lesssim 0.1\,M_\sun$, e.g., brown dwarfs.
As shown in Figure~\ref{time-pmass.eps}, some sink particles have 
$\sim 0.1 \, M_\sun$ when they are produced.  
If a finer resolution is adopted, the initial masses of the sink
particles could be lower. This could shift the turnover to lower
mass.

Figure~\ref{time-pmass.eps} shows the evolution of sink particle mass
for the four representative models.  As also shown in
Table~\ref{table:model-parameters}, 
the number of sink particles is larger in the high-density models
than in the low-density models.
The high-density models and the models with colliding flows take
a shorter time for the sink particles to reach $10\,M_\sun$ 
after the initial sink particle formation
than do the other models. 
However, the typical accretion rates (slopes of the lines) do not exhibit a significant difference
between models. 
Among all sink particles, 
those reaching $10\,M_\sun$, as indicated by 
the thick lines, show slightly higher accretion rates (steeper slopes) 
for the high-density and/or colliding models.

\begin{figure}
\epsscale{1.}
\plotone{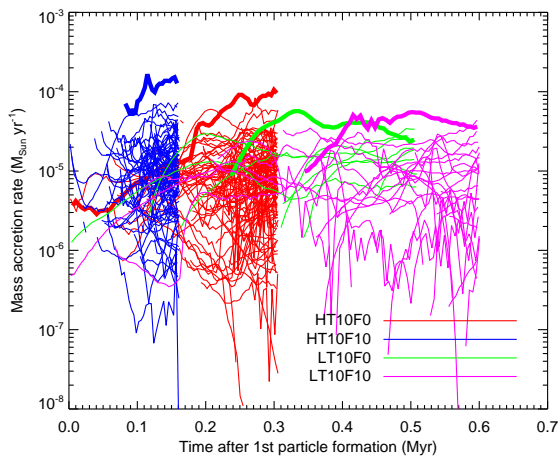}
\figcaption[]{ 
Mass accretion rates of sink particles as functions of time after the first sink
particle formation for models HT10F0, HT10F10, LT10F0, and LT10F10.
The thick lines show particles reaching $10\,M_\sun$.
\label{time-mdot.eps}
}
\end{figure}

Figure~\ref{time-mdot.eps} shows the accretion rates for the sink particles
explicitly. The accretion rates are spread over a wide range of 
$10^{-7}\,M_\sun \mathrm{yr}^{-1} \lesssim \dot{M} \lesssim 10^{-4}\,M_\sun \mathrm{yr}^{-1}$,
which is a common range for the models.
The accretion rate for each particle oscillates within this range.
This range includes the typical accretion rate for gas of 10~K, 
i.e., $c_s^3/G = 2 \times 10^{-6}\,M_\sun \mathrm{yr}^{-1}$, while 
the majority of sink particles exceed this value.
The most massive particle tends to have the highest accretion rate in each model
(see thick lines in Figure~\ref{time-pmass.eps} and Figure~\ref{time-mdot.eps}).
\revise{
  There is also a tendency that a model producing more stars has higher
  maximum mass of stars when observed at a given epoch.  For example,
  among the four models shown in Figure~\ref{time-pmass.eps}, 
  model HT10F10 has the highest maximum stellar mass, and 
  this model produces the most number of stars
  at 0.1 Myr after the first sink particle formation.
  This indicates that the maximum mass of stars correlates with the
  number of stars formed there.
}
A similar implication is proposed by the observational research of 
\citet{Dobashi01}, in which the maximum stellar luminosity of protostars
correlates with the mass of the parent clouds.
\revise{
Moreover, some other observations have shown that the number of protostars also correlates with
the mass of the parent cloud \citep{Dobashi96,Yonekura97,Kawamura98}.
When we simply assume that the stellar luminosity correlates
with the stellar mass, these observations also show that the maximum
stellar masss correlates with the number of stars in agreement with
our simulations.
}

\begin{figure}
\epsscale{1.}
\plotone{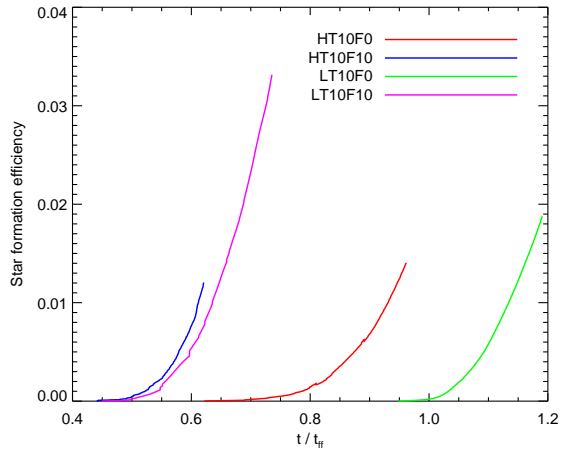}
\figcaption[]{ 
Star formation efficiency as functions of time 
for models HT10F0, HT10F10, LT10F0, and LT10F10.
\label{time-sfe.eps}
}
\end{figure}

Figure~\ref{time-sfe.eps} shows the star formation efficiency as a
function of the freefall time for each model.
The star formation efficiency is defined as the ratio of the mass of
all sink particles to the total mass of all particles and gas.
Star formation in models with colliding flows (models HT10F10
and LT10F10) begins in earlier stages than for models without colliding
flow (models HT10F0 and LT10F0). 
In comparing models with the same colliding flow, 
the high-density models tend to begin star formation earlier than do
the low-density models.
After star formation begins, the star formation
efficiencies increase with a similar trend in all models. 
This similar trend indicates that the high-density models have
higher accretion rates in total than do the low-density models because the
clouds of the high-density models are more massive than those of the
low-density models.
These high accretion rates are mainly due to the large number of sink
particles formed in the high-density models. 

\subsection{Velocity distribution for sink particles}
\label{sec:velocityDistribution}

\begin{figure*}
\epsscale{1.0}
\begin{center}
\plottwo{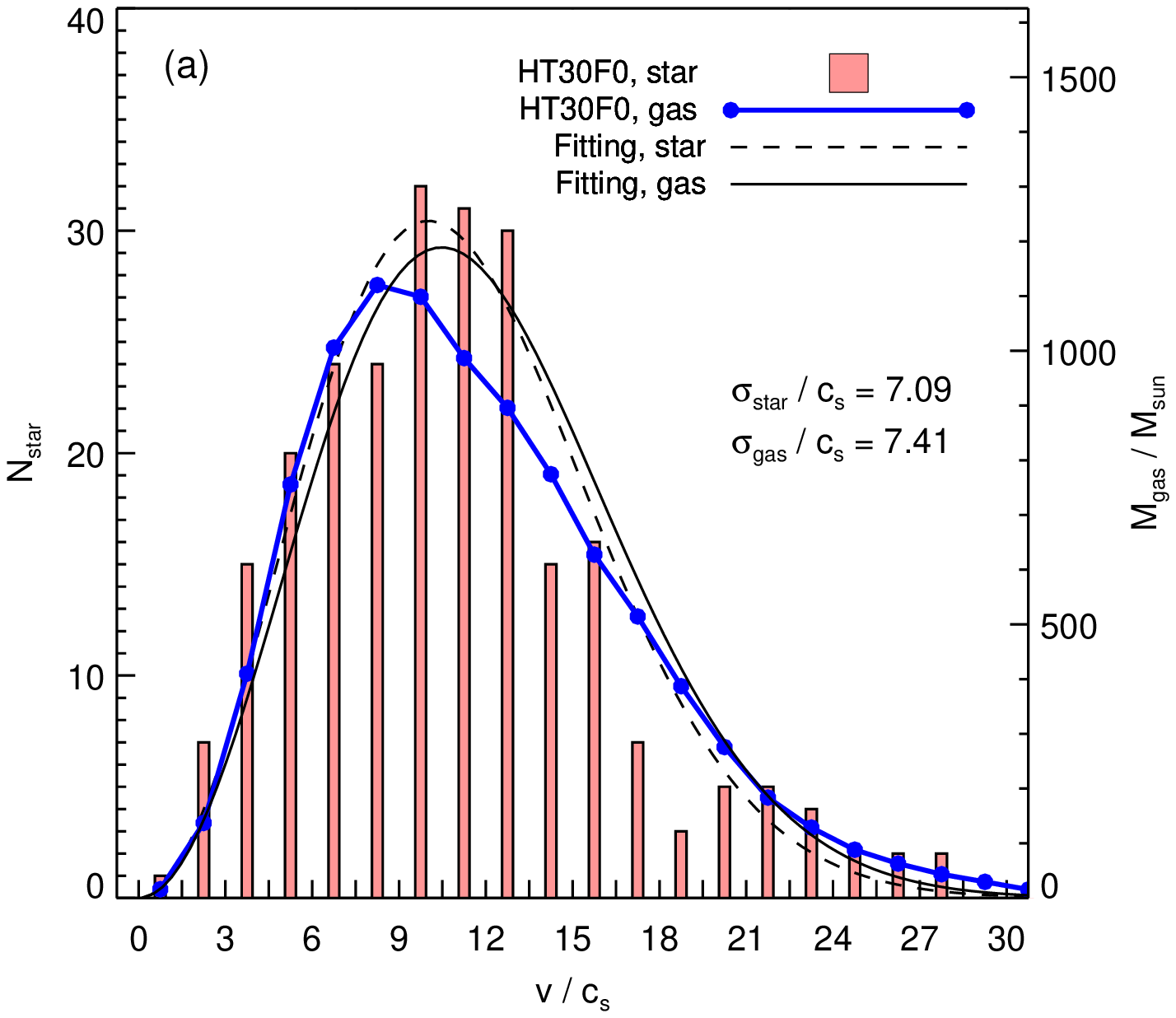}{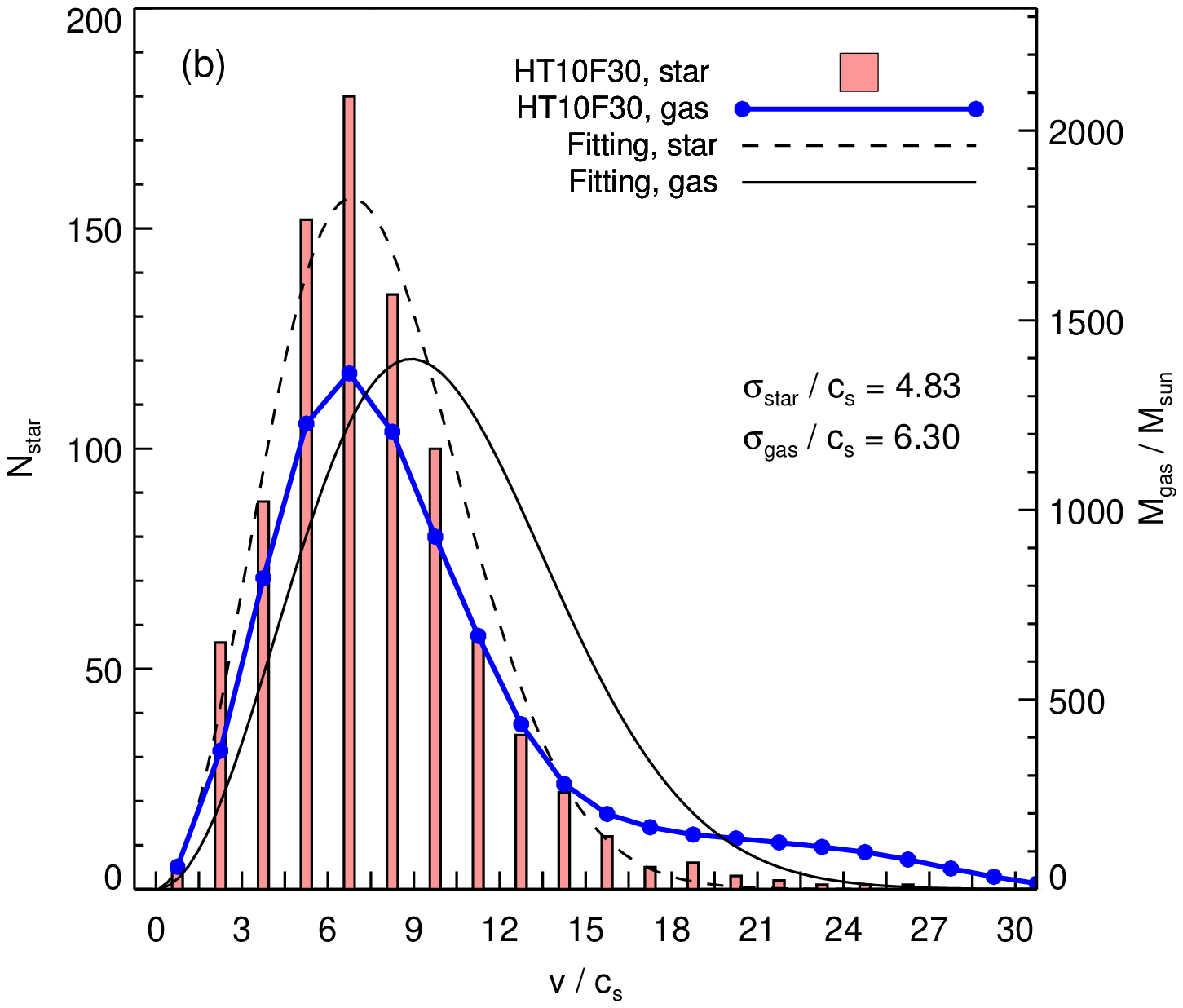}
\plottwo{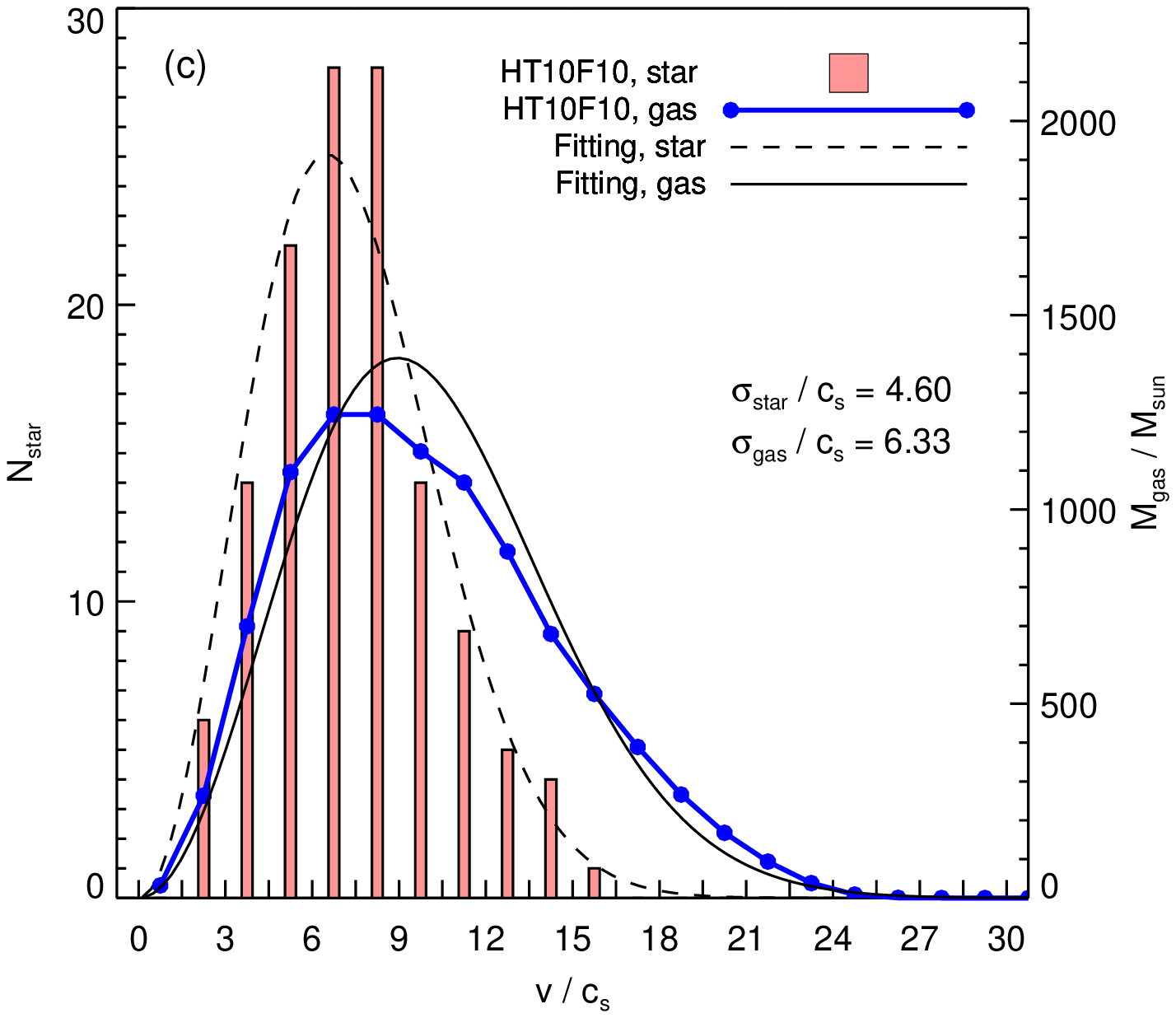}{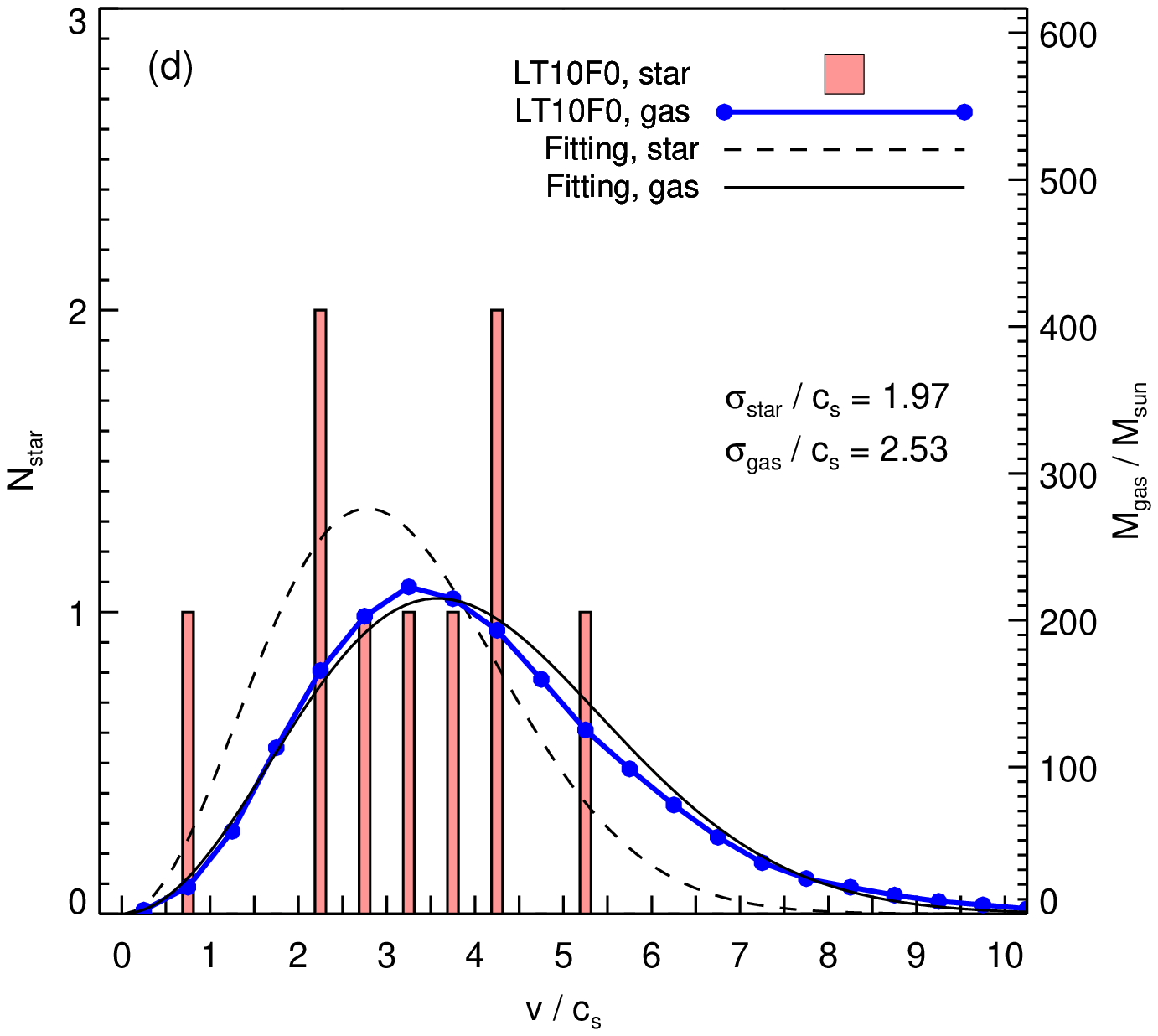}
\end{center}
\figcaption[]{ 
Histograms of sink particles (bar charts) and gas (blue lines with
filled circles) as functions of velocity $v/c_s^2$
for models 
(a) HT30F0, (b) HT10F30, (c) HT10F10, and (d) LT10F0
at the stage of $M_\mathrm{\star, max} = 10\,M_\sun$.  
The dashed and solid lines are fitting curves derived from the 
functions of $p(|\bmath{v}|; \sigma_\mathrm{star})$ and  $p(|\bmath{v}|; \sigma_\mathrm{gas})$ 
(see eq~(\ref{eq:gaussian-dist})).
\label{pv-gv-histogram.eps}
}
\end{figure*}

Sink particles are produced with initial velocities
so that the linear momenta of the material gas and the formed sink particles
are conserved. 
Sink particles change their velocities via 
gas accretion onto the sink particles and 
gravitational interaction with the gas and other particles.  
Figure~\ref{pv-gv-histogram.eps} shows the velocity distributions for 
sink particles in bar charts for the four representative models 
HT30F0, HT10F30, HT10F10, and LT10F0
at the stage of $M_\mathrm{\star, max} = 10\,M_\sun$.  
The high-density models have many sink particles, and each model exhibits a
profile of velocity distribution with a peak and tails at both low and high velocities.
In contrast, the low-density models do not have sufficient
numbers of sink particles to show such a clear profile.

The velocity distributions for gas are plotted for comparison in
Figure~\ref{pv-gv-histogram.eps}.  For the high-density models
HT30F0 and HT10F30, the velocity distributions for sink particles
are similar to those for gas; the peak positions and widths of the
profiles coincide between the sink particles and the gas.  For model
HT10F30, the velocity distribution for gas has a wing at high
velocities because of the accretion flow onto the sheet cloud.  The
accretion flow does not considerably affect the velocities of the sink
particles because the accretion flow has low density and the sink
particles exist around the high-density sheet clouds.
Model HT10F10 also has an accretion flow onto the sheet, but
the flow is denser and slower than that for model HT10F30 
(see Figure~\ref{colden_F_HT10.eps} for comparison of
the densities of accretion flows between these models).
The profile of the velocity distribution for gas therefore 
appears to shift
to the high velocities with respect to the velocity distribution for the sink particles.

For all low-density models, 
the velocity distributions for sink particles roughly agree with those for gas,
even though the number of sink particles is small.
Note that sink particle formation occurs after the 
sheet caused by the colliding flow disappears for the low-density models.
Even for the colliding models, the velocities of sink
particles follow the velocity distribution for gas.

The fitting curves for both sink particles and gas are also plotted 
in Figure~\ref{pv-gv-histogram.eps} under the assumption that the 
velocities of sink particles and gas follow a Gaussian
distribution.
When each velocity component, $(v_x, v_y, v_z)$, follows a Gaussian
distribution with a velocity dispersion, $\sigma$, 
the distribution for the absolute value of the velocity $|\bmath{v}|$
is given by 
\begin{equation}
p(|\bmath{v}|; \sigma) = \frac{2 |\bmath{v}|^2}{(2 \pi)^{1/2} \sigma^3} 
\exp\left(-\frac{ |\bmath{v}|^2 }{2 \sigma^2}\right).
\label{eq:gaussian-dist}
\end{equation}
The velocity dispersion for the sink particles is evaluated by 
\begin{equation}
\sigma_\mathrm{star} = \left(
\frac{1}{ 3N_\star}
 \sum_{i=1}^{N_\star}  |\bmath{v}_i|^2
\right)^{1/2}, 
\label{eq:sigma_star}
\end{equation}
and that for gas is evaluated by using a mass-weighted integral,
\begin{equation}
\sigma_\mathrm{gas} = \left(
\frac{\displaystyle \int \rho(\bmath{r}) |\bmath{v}(\bmath{r})|^2 dV}
{\displaystyle 3 \int \rho(\bmath{r}) dV}
\right)^{1/2}.
\label{eq:sigma_gas}
\end{equation}
The numerical factor of 3 in Equations~(\ref{eq:sigma_star}) and
(\ref{eq:sigma_gas}) comes from the number of spatial dimensions.

The fitting curves indicate that the velocities of sink particles
and gas are well described by a Gaussian distribution for 
model HT30F0.   The velocity
dispersion for sink particles is roughly the same as that for gas
($\sigma_\mathrm{star} \simeq \sigma_\mathrm{gas}$).
Such a tendency is confirmed in all the high-density models without 
colliding flow. 
In the high-density models with colliding flow, gas has
a significantly larger velocity dispersion than that for sink particles
($\sigma_\mathrm{gas} \gtrsim \sigma_\mathrm{star}$) because of
the high-velocity wing due to the colliding flow.
Due to the large velocity dispersion, the fitting curve for gas shifts
to high velocities with respect to sink particles.
In these models we also confirmed that, for sink particles,
the velocity dispersion for $v_x$ tends to be smaller than those for $v_y$ and
$v_z$. The dispersion difference indicates that the gravity of the sheet decelerates the motion
of the sink particles along the $x$-direction, which is the direction normal to
the sheet.

Sink particles and gas change their velocity distributions as
time proceeds.  
For model HT30F0, 
a considerable decrease in the velocity dispersion for gas is seen
in Figure~\ref{logugr_plot_hl_lin.eps},
even after formation of the first sink particle.
In
this model, the velocity dispersion for gas decreases:
$\sigma_\mathrm{gas}/c_s = 10.76$ and $7.41$  
at the stages of $M_\mathrm{\star, max} =1\,M_\sun$ and $10\,M_\sun$, respectively.
The velocity dispersion for sink particles also decreases: 
$\sigma_\mathrm{star}/c_s = 9.09$ and $7.09$ 
at the stages of 
$M_\mathrm{\star, max} = 1\,M_\sun$ and $10\,M_\sun$, respectively.
This indicates that stars with a velocity dispersion similar to
that for gas are produced continuously. 
On the other hand, for model HT10F0, the velocity dispersion for gas
decreases slightly, as seen in Figure~\ref{logugr_plot_hl_lin.eps}, and 
both $\sigma_\mathrm{star}$ and $\sigma_\mathrm{gas}$ are roughly constant.

\subsection{Channel maps}

Reconstruction of channel maps by using the data of the numerical
simulations is useful for comparison of observations and
numerical simulations, as demonstrated by \citet{Dobashi14}.
The channel maps here are reconstructed for four models:
HT10F0, HT10F10, LT10F0, and LT10F10.
In reconstructing a channel map, gas is assumed to be optical thin and 
both the non-thermal and thermal velocity components are considered;
each numerical cell has a velocity along the line of sight, $v_\mathrm{los}$, with
a Gaussian profile, which is broadened by the sound speed, and 
the central velocity of the Gaussian is equal to the bulk velocity.
The $x$-direction is adopted as the line of sight,
which is parallel to the direction of the colliding flow.

\begin{figure*}
\epsscale{1.0}
\begin{center}
\plottwo{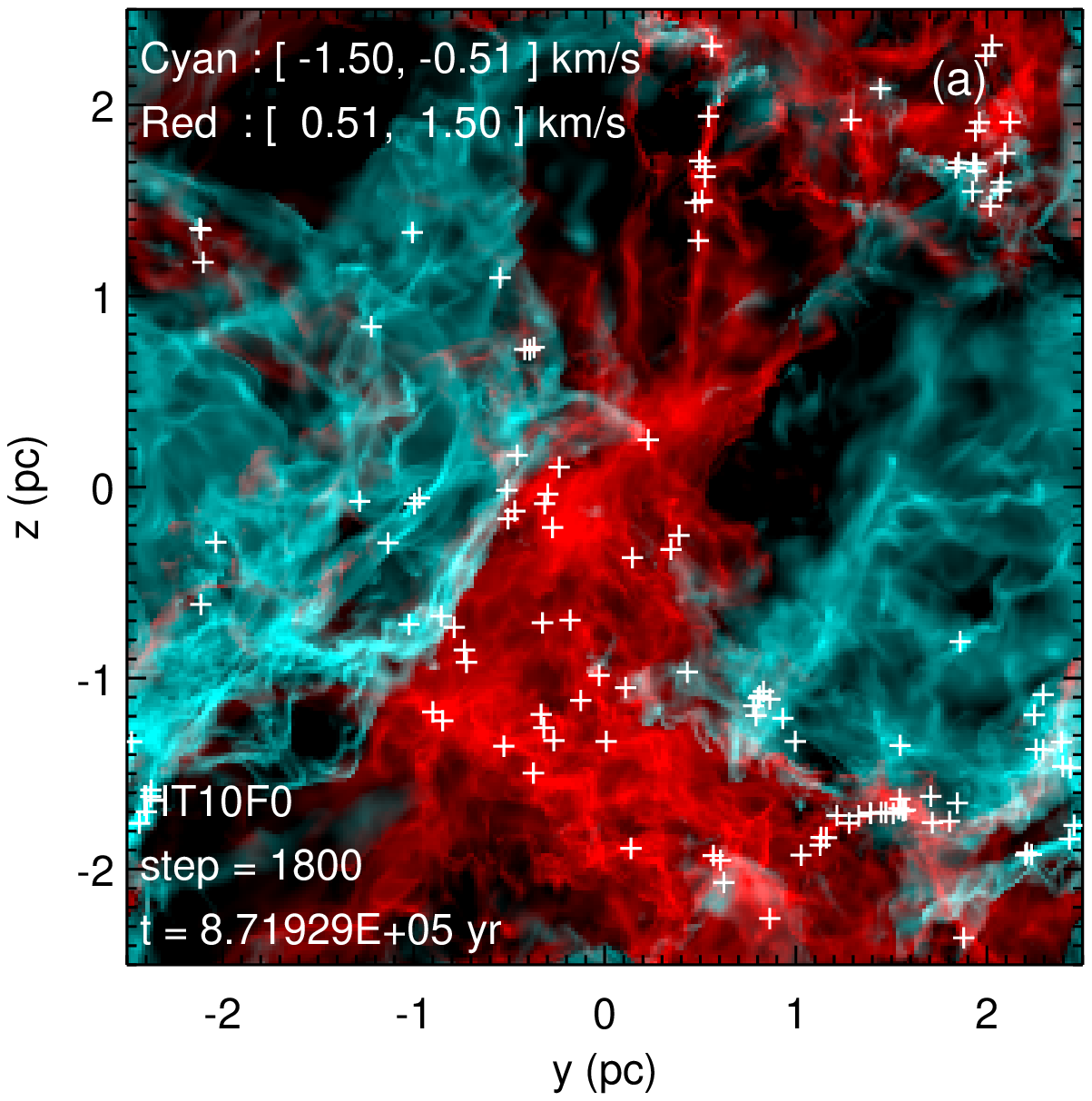}{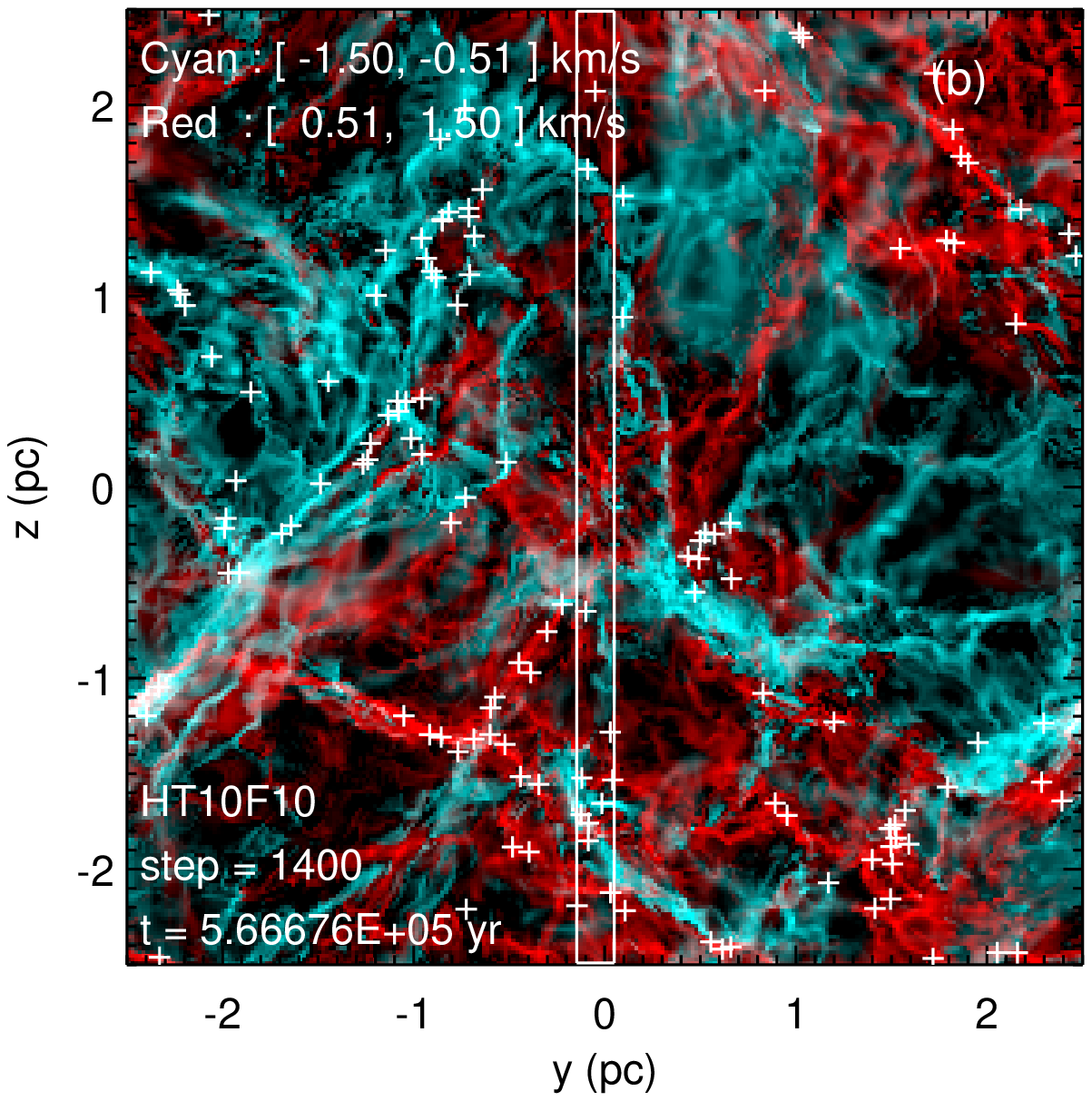}
\plottwo{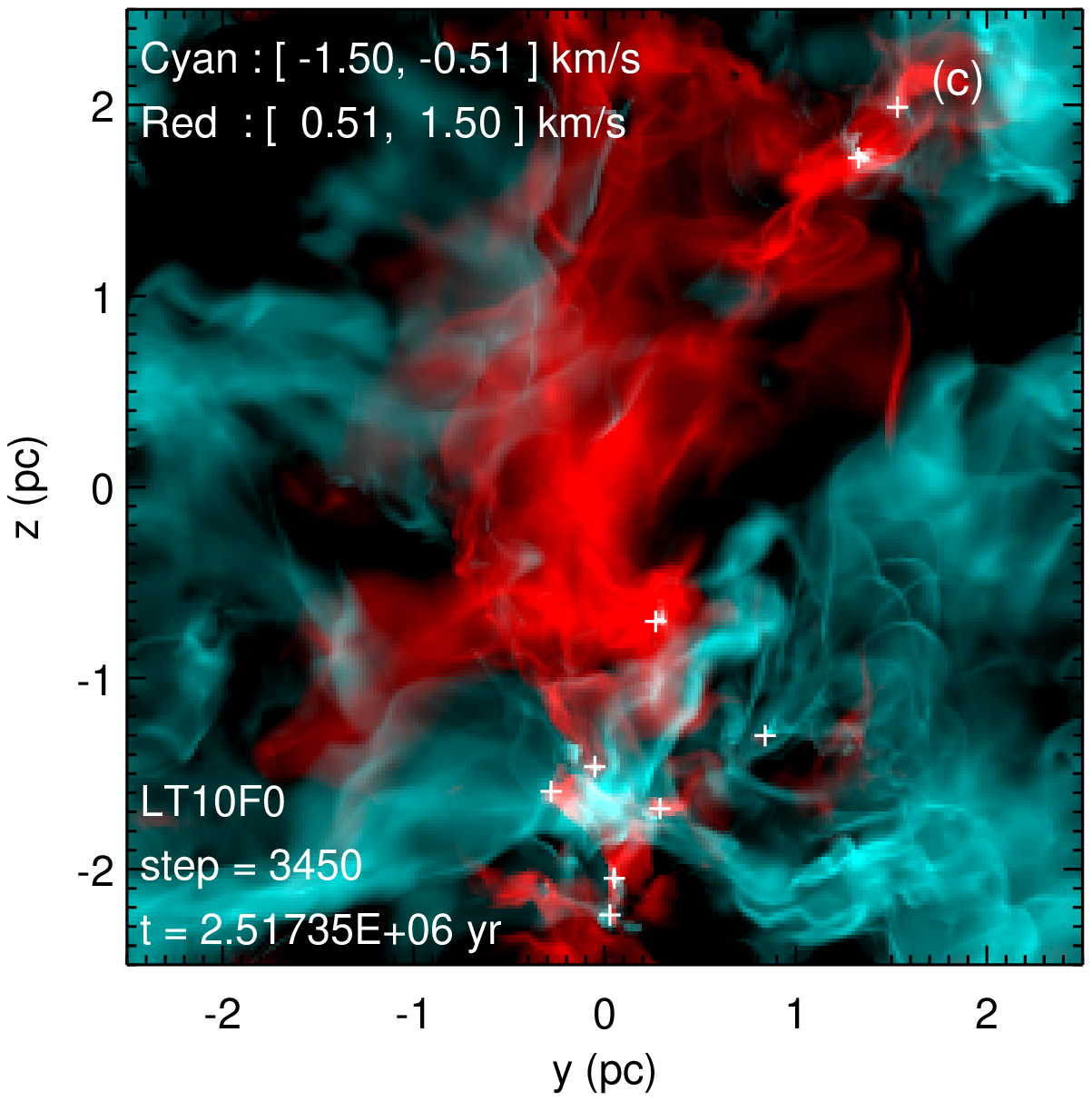}{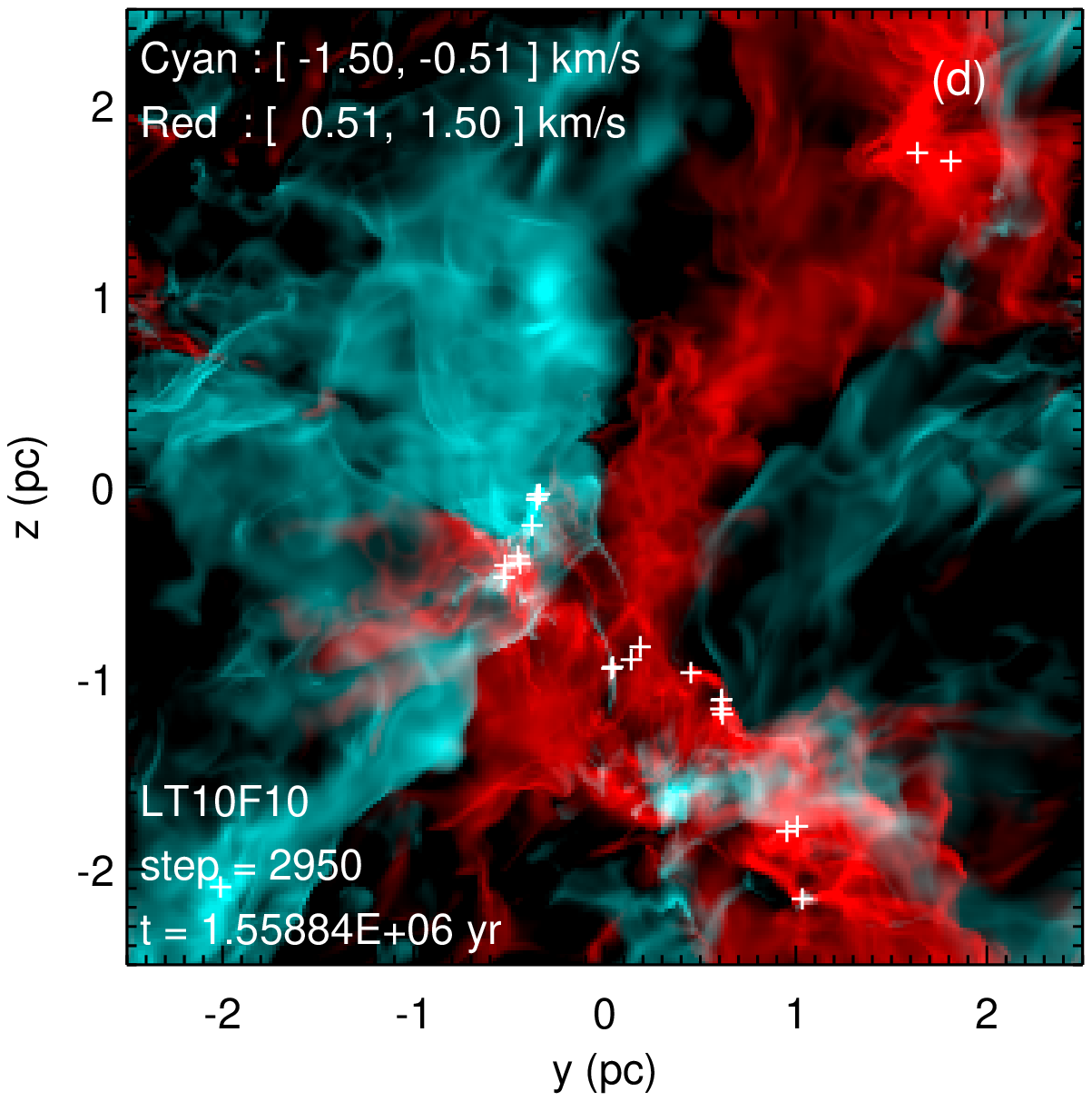}
\end{center}
\figcaption[]{ 
Composite maps of different velocity channels for models
(a) HT10F0, (b) HT10F10,
(c) LT10F0, and (d) LT10F10 at the stages of 
$M_\mathrm{\star, max} = 10\,M_\sun$.
Two velocity channels of $-1.50\,\mathrm{km\,s^{-1}}<v_\mathrm{los}<-0.51\,\mathrm{km\,s^{-1}}$
(cyan) and $0.51\,\mathrm{km\,s^{-1}}<v_\mathrm{los}<1.50\,\mathrm{km\,s^{-1}}$ (red) are
overlaid.
Note that red and cyan are complementary colors.
The regions where the gas exists in both channels are shown in white, 
those having gas in either channel are shown in red or cyan,
and those without gas in both channels are shown in black.
The plus symbols ($+$) denote the positions of the sink particles.
The slit in panel (b) is for the position-velocity diagram shown in Figure~\ref{pv-diagram.eps}.
\label{channelMap_hole.eps}
}
\end{figure*}

Figure~\ref{channelMap_hole.eps} shows 
composite maps for the two velocity channels.
In each composite map, two representative velocity channels of 
$-1.50~\mathrm{km~s}^{-1}\leq v_\mathrm{lols} \leq -0.51~\mathrm{km~s}^{-1}$ and 
$0.51~\mathrm{km~s}^{-1} \leq v_\mathrm{lols} \leq 1.50~\mathrm{km~s}^{-1}$ are 
shown in cyan and red colors, respectively. 
Components existing in both velocity channels are expressed in white because 
cyan and red are complementary colors.

The most prominent features are shown in model HT10F10
(Figure~\ref{channelMap_hole.eps}(b)),
where red and cyan filaments are tangled up with each other on a 0.1~pc scale. 
Moreover, the red and cyan filaments show an anti-correlated distribution 
because of the collision of filaments, as shown by
\citet{Dobashi14}. 
\revise{Similar anti-correlation between the different velocity components was
also reported by \citet{Shimoikura13}.}
A considerable number of sink particles exist near the region of collisional interaction, and
this indicates that the collision of filaments induces star formation.
Stars are also formed along the dense filaments, e.g., the red
filament running from $(y, z) = (-0.7, -1.3)~\mathrm{pc}$ to 
$(-0.2, -0.6)~\mathrm{pc}$.

In contrast, models HT10F0, LT10F0, and LT10F10 exhibit
coarser patterns of red and cyan colors on a parsec scale
(Figures~\ref{channelMap_hole.eps}(a), (c), and (d)). 
These patterns indicate that 
velocity shear on the cloud scale exists. 
This is attributed to turbulence, which has a larger kinetic energy in
a longer wavelength mode.  Therefore, the cloud-scale gas motion should
dominate over short wavelength modes when considering turbulence with 
the power spectrum examined here.
We confirmed that similar anti-correlations on the cloud scale are also 
shown irrespective of the choice of line of sight for the models
without colliding flows.

The low-density colliding model LT10F10 has a sheet cloud caused by
the colliding flow before sink particles form.  The sheet cloud
disappears at the stage shown in Figure~\ref{channelMap_hole.eps}(d).
When the sheet cloud exists, the channel map shows the
anti-correlation pattern on a sub-parsec scale, as also seen for model
LT10F10.  However, the contrast for the anti-correlation pattern is
much weaker than that for the high-density model HT10F10 because
the filaments in the low-density model have a low density contrast.

\begin{figure*}
\epsscale{1.0}
\plotone{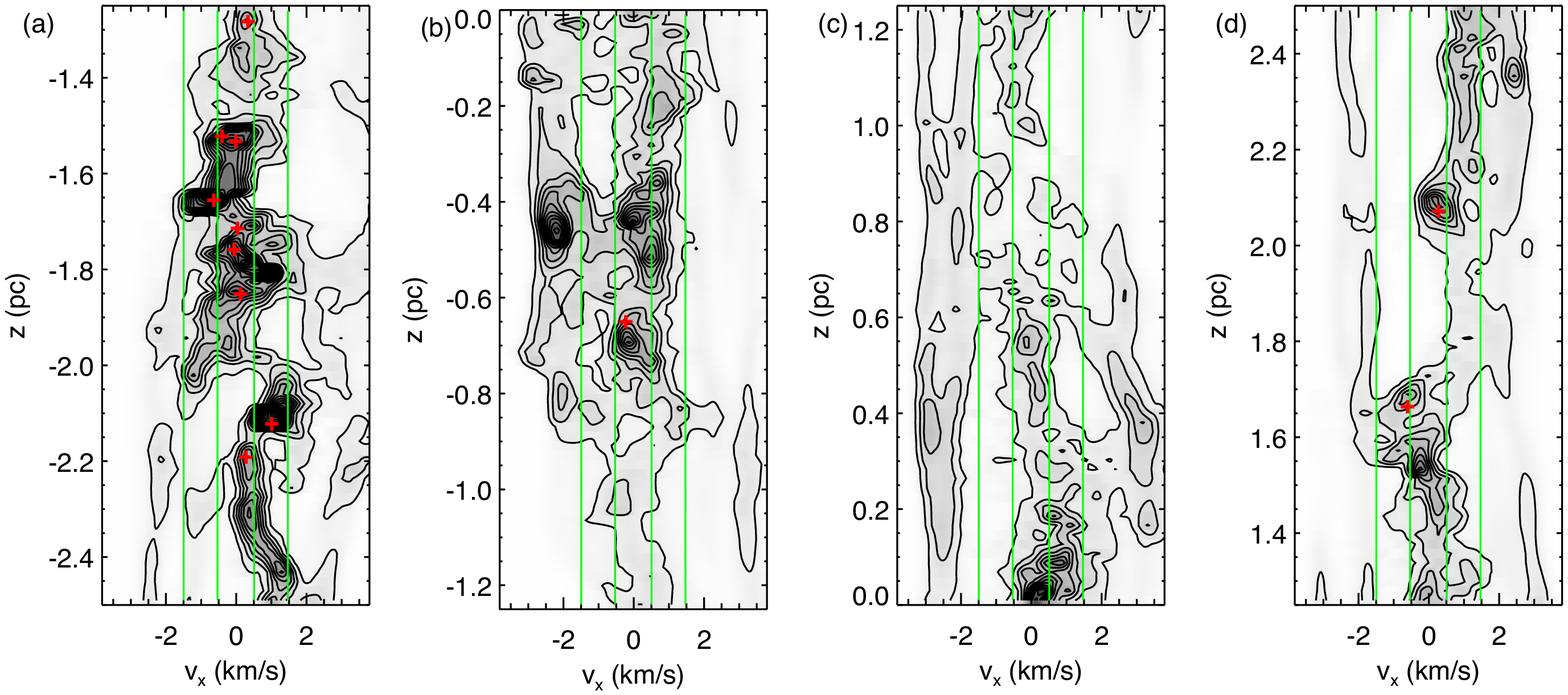}
\figcaption{
Position-velocity diagram for model HT10F10 at the stage of  
$M_\mathrm{\star, max} = 10\,M_\sun$.
The long slit region indicated in Figure~\ref{channelMap_hole.eps}(b) is
shown in the four panels separately according to the $z$-coordinates.
The contours and gray scales show the gas density.
The lowest contours and contour intervals are 
$2.5 \times 10^{38}\,\mathrm{cm}^{-1}(\mathrm{km\,s})^{-1}$.
The red plus symbols ($+$) denote the locations of the sink particles in the
position-velocity plane.
The vertical green lines show the boundaries of the two velocity ranges
shown in Figure~\ref{channelMap_hole.eps}.
\label{pv-diagram.eps}
}
\end{figure*}

Figure~\ref{pv-diagram.eps} shows a position-velocity (PV) diagram for
examining the slit region in Figure~\ref{channelMap_hole.eps}(b).  
Every sink particle is associated with a dense gas component in
the PV diagram.  Figure~\ref{pv-diagram.eps}(a) shows the PV
diagram for the region of $-2.5\,\mathrm{pc} \le z \le -1.25\,\mathrm{pc}$, 
where the interaction between clouds with different velocities induces
the star formation shown in Figure~\ref{channelMap_hole.eps}(b).
The PV diagram clearly shows that
dense gas components pass through each other with supersonic
velocities in the $x$-direction (line of sight).

In Figure~\ref{pv-diagram.eps}(b), 
two dense components with different velocities
exist in $-0.5\,\mathrm{pc} \lesssim z \lesssim -0.4\,\mathrm{pc}$
without an associated sink particle. This figure shows that 
these components have not yet collided.
The components are shown in white in the composite map of 
Figure~\ref{channelMap_hole.eps}(b).
Indeed, we confirmed that these components collide 0.2~Myr
later, and star formation is induced.

The regions shown in Figure~\ref{pv-diagram.eps}(c) and \ref{pv-diagram.eps}(d)
have only two sink particles, and active star formation is not induced
there.
In this region, anti-correlation between the dense cyan and the thin red components is observed
at $z \simeq 1.7\,\mathrm{pc}$ and a sink particle is associated
(Figure~\ref{channelMap_hole.eps}(b)).
The PV diagram suggests that the sink particle is associated with the cyan component,
and it interacts with the ambient gas 
with low density (Figure~\ref{pv-diagram.eps}(d)).

\section{Discussion}
\label{sec:discussion}

\subsection{Comparison with observations}

\subsubsection{Probability distribution functions}
\label{sec:discussion:PDF}
All the models without the colliding flow (${\cal M}_f = 0$) exhibit
PDFs that can be fitted by lognormal functions.  The models with
weak turbulence tend to show a power-law tail at high density
(Figures~\ref{pdf_plot_paper_M.eps} and \ref{pdf_plot_nsteps.eps}).  

\citet{Kainulainen09} showed that PDFs for most molecular clouds based
on the 2MASS data are well-fitted by lognormal functions at low column
densities, and active star-forming clouds always have prominent
non-lognormal wings or power-law wings.
Although they did not mention power indexes of the power-law wings explicitly, 
the power indexes appear in ranges from $-3$ to $-2$, according to their
figures.
Recently, \citet{Lombardi14} combined Plank dust-emission maps,
{\it Herschel} dust-emission maps, and 2MASS NIR dust-extinction maps for
the Orion complex. They showed that the PDFs for Orion A and B can be
fitted by a simple power law with indexes of $-3$.
The power-law wings obtained by observations 
are in agreement with those for our numerical simulations,
which have power indexes ranging from $-3$ to $-2$ (see Section~\ref{sec:PDF}).

\citet{Kainulainen09} reported that half of the clouds examined in
their paper showed non-lognormal features of PDFs at low column
densities. This feature is reproduced in our simulations.
According to the simulations examined here, the PDF at
low density oscillates considerably around a lognormal function, 
and the PDF obtained from a snapshot 
of a simulation should deviate from a lognormal function.

PDFs can be affected by the environment of the clouds, as demonstrated by
the models with colliding flows (${\cal M}_f \neq 0$).
According to \citet{Lombardi06}, the PDF for the Pipe molecular clouds 
is poorly fitted by a single lognormal function.
The PDF has a dip near its peak that resembles
the PDF for model HT10F10 shown in Figure~\ref{pdf_plot_nsteps.eps}(b). 
This suggests that the PDF for the Pipe
nebula is influenced by external factors.
This cloud can be affected by the wind from
$\tau$~Sco, a B0 star, as suggested by \citet{Onishi99}.
In similar cases, 
PDFs deviating from a lognormal function were reported by 
\citet{Froebric10} in Cepheus, Monoceros, and
Ophiuchus clouds.

\subsubsection{Velocity distributions}

In our simulations, the velocity distributions for the sink particles
are similar to those for gas.
Our results agree with 
the kinematic studies of the T-Tarui stars, which were investigated based on
their radial velocities and proper motions in several star-forming regions
\citep[e.g.,][]{Herbig77,Jones79,Hartmann86,Dubath96,Makarov07}.
These studies reported that the mean stellar velocities are roughly the same as the local
gas velocities. In addition, the one-dimensional velocity dispersions obtained by
these studies ($\simeq 1-2\,\mathrm{km\,s}^{-1}$) are consistent with the
line widths of the molecular lines.  
Similar results were obtained for protostars by \citet{Covey06}.
\citet{Furesz08} found that the radial velocities of stars appear to
correlate strongly with the radial velocity of the molecular gas cloud
for the Orion Nebula Cluster (ONC).
\citet{Tobin09} also investigated the kinematics of ONC and 
showed that the stellar velocities tend to follow the local
gas velocities.
These studies as well as our results support the paradigm of dynamical star formation
proposed by \citet{Hartmann01}, 
\revise{
where rapid star formation is induced by turbulence and/or
external flows, and the formed stars are scattered
at the local gas velocity as shown in our simulations.  
In this paradigm, star formation
should be terminated by rapid dispersal of gas due to stellar feedback, which is not
taken into account in our simulations.  Therefore in our simulations,
the stars continue to accrete gas, and star formation efficiency
becomes high at the late stages of the cloud evolution. 
}

\subsection{Comparison with other simulations}

As shown in the analysis of the velocity power in
Section~\ref{sec:compressibility_of_velocity_fields}, 
non-gravitational non-colliding model HT10F0wog
keeps $E_\mathrm{long}/E_\mathrm{tot}=0.15-0.2$.
This ratio agrees with that obtained by \citet{Federrath10},
who found that non-gravitational isothermal driven
turbulence with ${\cal M}_t \simeq 5.5$ has a ratio of $E_\mathrm{long} / E_\mathrm{tot} \simeq
0.2$ when the driving force is 1/3 or
less of the longitudinal mode, i.e.,
$F_\mathrm{long}/F_\mathrm{tot} \le 1/3$ (see their Figure~8).  
The ratio of $E_\mathrm{long}/E_\mathrm{tot} \simeq 1/3$ is expected
from geometrical considerations \citep[e.g.,][]{Nordlund03}.
\citet{Federrath10} showed that 
the velocity Fourier spectra shows
$E_\mathrm{long}(k)/E_\mathrm{tot}(k) \simeq 1/3$ in the inertia range
of $k$ for the solenoidal driving force.
For the decaying turbulence presented here, 
the ratio of $E_\mathrm{long}(k)/E_\mathrm{tot}(k)$ in the inertia range tends to decrease
from 0.4 to 0.25 as time proceeds, and the ratio agrees
with the case of the turbulence driven by solenoidal force. 

In the case with colliding flow,
model HT10F10wog shows a higher value of 
$E_\mathrm{long}/E_\mathrm{tot}$ in the early stage. This result
suggests that the colliding flow provides more compressibility.
The compressible flow shows a rapid decay, and 
$E_\mathrm{long}/E_\mathrm{tot} = 0.2 - 0.25$ in 
a long period of $(2-5) t_\mathrm{cross}$, 
\revise{which are slightly higher values than those in the
  non-colliding model.
Compared between models HT10F0wog and HT10F10wog, 
$E_\mathrm{trans}$ shows almost the same values during the decay
of turbulence, while
$E_\mathrm{long}$ exhibits slightly higher values in model HT10F10wog
than in model HT10F0wog after the rapid decay of $E_\mathrm{long}$ in
the early stages of model HT10F10wog.  The slightly higher value of
$E_\mathrm{long}/E_\mathrm{tot}$ in the colliding model is due to the
slightly higher value of $E_\mathrm{long}$.
\citet{Federrath10} showed that
non-gravitational turbulence shows $E_\mathrm{long}/E_\mathrm{tot} =
0.6$ when the driving force is purely compressive.  This ratio is
higher than the colliding model presented here. 
This is because compressibility is continuously supplied by the
driving force in the case of \citet{Federrath10}, while it is provided
by the colliding flow only in the initial stage in our model.}

In our simulations, 
the gravitational collapse leading to star formation occurs
locally compared with the cloud scale, and 
the density structure on the cloud scale is mainly a result of 
the interaction of turbulent flows.
However, self-gravity enhances the compressibility of the velocity field 
in the whole cloud.
For example, models HT3F0 and LT3F0 exhibit 
$E_\mathrm{long}/E_\mathrm{tot} \simeq 0.6$ at the last stages.
Such a high compressibility of the velocity field requires an extreme
driving force of $F_\mathrm{long}/F_\mathrm{tot} \gtrsim 0.9$
in the case of non-gravitational turbulence \citep{Federrath10}.

\citet{Hansen11} examined the decay of anisotropic turbulence and showed
that the decay rate of the turbulence depends on the crossing time of
the isotropic component only, and the decay rate is 
much lower for anisotropic turbulence than for isotropic turbulence.
For their highly anisotropic turbulence, $\langle |\nabla \times \bmath{v}|^2\rangle$
is much larger than $\langle |\nabla \cdot \bmath{v}|^2\rangle$, and so it seems
that the solenoidal flow dominates over
the compressive flow.
In our case, 
models with colliding flows can be considered as anisotropic
turbulent models, in which the colliding flow corresponds to a Fourier
component of $k_x=1$ in the compressive mode.
In contrast to the anisotropic models of \citet{Hansen11},
the velocity dispersion due to the colliding flow decays
faster than that for the isotropic turbulence (see Figure~\ref{logugr_plot_hl_lin.eps}(b)).
The anisotropic turbulence therefore decays faster than
the isotropic turbulence because anisotropy is responsible for the
compressible mode.

\subsection{Effects of magnetic field}

The magnetic field is ignored in this paper, although it is known that the
magnetic field plays an
important role in star formation.  

Many simulations have been performed
for turbulent clouds while taking into account magnetic
fields with and without self-gravity \citep[e.g.,][]{Maclow99,Ostriker01}.  
When considering a relatively strong magnetic field (plasma beta
$\simeq 0.2$), a turbulent cloud tends to collapse along
the magnetic field lines, and then a sheet structure oriented perpendicular
to the magnetic field tends to form \citep[e.g.,][]{Nakamura08}.
Even when strong turbulence of ${\cal M}\sim 10$ is imposed on the
initial stage, rapid decay of the turbulence leads to formation of a sheet structure
perpendicular to the magnetic field. 

In the magnetically critical case, 
the magnetic field strength is given by 
$B_\mathrm{cr} = 2\pi G^{1/2} \Sigma$ \citep{Nakano78,Tomisaka88},
where $\Sigma$ denotes the column density for a typical scale and 
is estimated here as $\Sigma = \rho_0 \lambda_J$.
The magnetic energy of the critical field strength is therefore given by 
\revise{
$B_\mathrm{cr}^2/(8\pi) = (\pi^2/2) \rho_0 c_s^2 = 4.9 \rho_0 c_s^2$.
}
This indicates that a flow faster than 
\revise{ $\sim 3 c_s$  }
dominates over the
magnetic field, and a sheet forms due to compression by the flow
rather than the magnetic field. 

\revise{
\citet{Chen14} performed MHD simulations of turbulent clouds with
a colliding flow, taking into account ambipolar diffusion. 
Their model settings are similar to ours except for the magnetic field; 
they assumed an initial number density of $n = 10^3\,\mathrm{cm}^{-3}$
and a colliding flow of Mach~10. 
Because of the high Mach number of the colliding flow, a sheet forms
due to the compression of the flow, and the gas and the magnetic field are
accumulated into the sheet.
The sheet is perturbed little compared to our cases
because they assume relatively weak turbulence.
The evolution of the sheet cloud is highly dependent on the ionization rate and
the direction of the magnetic field.  
The models with a low ionization rate exhibit
several features similar to the hydrodynamical models, e.g., 
the post-shock densities, collapse times, masses of cores.
Thus, several features of our hydrodynamical models may hold when
our models are extended to strongly diffusive MHD.
}

\section{Summary}
\label{sec:summary}
The formation of dense molecular clouds and star formation
are investigated by high-resolution numerical simulations that 
take into account turbulence and colliding flow.
The main outcomes are summarized as follows:
\begin{enumerate}
\item The interaction of shock waves due to turbulence produces
  filamentary structures.  Some of the filaments have sufficient
  density to undergo gravitational collapse to form stars. 
  When the colliding flow has a velocity equal to
  or larger than the rms turbulent velocity (${\cal M}_f \ge {\cal M}_t$), 
  the filaments are accumulated into a sheet cloud. 
\item Colliding flow, strong turbulence, and an initial high
  density promote active star formation.  All of these contribute to
  the formation of dense and thin filaments, which are unstable against 
  gravitational collapse. 
\item The turbulence decays in the crossing timescale of turbulent
  flow, while the colliding flow decays rapidly.
  The compressibility of the velocity field has a fiducial value of
  $E_\mathrm{long}/E_\mathrm{tot} \simeq 0.15-0.2$, which is consistent with
  the case of non-gravitational isothermal driven turbulence.
  Self-gravity increases the compressibility of the velocity
  field considerably.  The colliding flow increases the compressibility only in the early stages.
\item PDFs can be well fitted by lognormal functions for highly
  turbulent models
  without colliding flow, but the PDFs deviate from the
  lognormal feature at the peaks for models with strong colliding flows. 
  For weak turbulent models, the PDFs tend to exhibit power-law
  features at high density in the later stages.  This is 
  in agreement with recent observations of star-forming clouds. 
\item The high-density models produce a sufficient number of
  stars for examining a mass function.
  The histograms of the stellar masses can be roughly fitted by the
  classical IMF of \citet{Salpeter55},
  $N_\mathrm{star}(M_\mathrm{star}) \propto M_\mathrm{star}^{-1.35}$
  for the high-density models except for the extremely strong
  flow model HT10F30.  
\item The stellar velocities are distributed in agreement with 
  the velocity distribution for the gas of the parent clouds.  
  Dispersions of the stellar velocity are similar to
  the velocity dispersion for gas ($\sigma_\mathrm{star} \simeq
  \sigma_\mathrm{gas}$) for models without colliding flow.
  Colliding flow can increase the velocity dispersion for gas, such
  that $\sigma_\mathrm{gas} \gtrsim \sigma_\mathrm{star}$.
\item  Composite channel maps in complementary colors display
  the interaction of the gas of different velocity channels.  The maps
  illustrate characteristics of the velocity distribution due to 
  turbulence and collisions of filaments.
  The collision of filaments is observed as 
  an anti-correlated distribution of thin filaments between
  the different velocity channels.
\end{enumerate}

\acknowledgments 

Numerical computations were carried out on the Cray XC30 at the Center for
Computational Astrophysics, CfCA, of the National Astronomical Observatory
of Japan.
This research was supported by
JSPS KAKENHI Grant Numbers
22340040, 
23540270, 
24244017, 
26287030, 
26400233, 
26350186,
26610045.

\begin{appendix}
\section{Method of Sink Particles}
\label{sec:sink_particle}
We implemented sink particles in our AMR code, SFUMATO
\citep{Matsumoto07}.  The sink particles are Lagrangian particles,
which move on the numerical grid for hydrodynamics.  The particles interact with
gas through gravity and gas accretion.  A sink particle has
the properties of mass, position, velocity, and spin angular momentum.  In
this section we show the implementation of the sink particle method
adopted in this paper.

\subsection{Creation of Sink Particles}
When a
high-density portion  undergoes gravitational collapse, a sink particle
is created there.  Sink particle creation is implemented mainly
according to \cite{Federrath10a}; the conditions for particle
creation are that (1) the density is higher than the threshold density,
$\rho_\mathrm{sink}$, at the position of the new particle, (2) the
gravitational potential takes the local minimum there, (3) the
divergence of the velocity is negative ($\nabla \cdot \bmath{v} < 0$),
(4) all the eigenvalues of the 
symmetric parts of the velocity gradient tensor $\nabla \bmath{v}$ are
negative, (5) the total energy of the gas within the sink radius
$r_\mathrm{sink}$ is negative ($E_\mathrm{th} + E_\mathrm{kin} +
E_\mathrm{grav} < 0$), and this indicates that the gas is gravitationally
bound there.  In addition, sink particle creation is forbidden at a
position within a distance of $2 r_\mathrm{sink}$ from other sink
particles in order to avoid an overlap of sink regions,
where $r_\mathrm{sink}$ denotes a sink radius, which is set at 
$r_\mathrm{sink} = 4 \Delta x$ in the finest grid level \citep[c.f.,][]{Krumholz04}.
These
conditions are checked at every hydrodynamical timestep, and a sink
particle is created when the conditions are satisfied.
The merger of sink particles can also be implemented, but this function is
switched off in the simulations presented in this paper.

\subsection{Gas Accretion onto Sink Particles}

The method for accretion is the same as that of \citet{Machida10}. A sink
particle accretes gas that is located within the distance of the sink
radius, $ r_\mathrm{sink}$, from the sink particle.  The accretion
onto sink particles is performed at each  hydrodynamical timestep.  The
mass accreted onto the $i$-th sink particle is given by
\begin{equation}
\dot{M} \Delta t = 
\int_{|\bmath{r} - \bmath{r}_i| < r_\mathrm{sink}}\Delta \rho(\bmath{r}) dV,
\label{eq:sink_mass_accretion}
\end{equation}
\begin{equation}
\Delta \rho(\bmath{r}) = \max\left[ \rho(\bmath{r}) - \rho_\mathrm{sink}, 0\right],
\label{eq:sink_mass_accretion_density}
\end{equation}
where $\dot{M}$ and $\Delta t$ 
denote the mass accretion rate and the hydrodynamical timestep, respectively.
The integration of Equation~(\ref{eq:sink_mass_accretion})
is a volume
integration within a sphere with a radius of $r_\mathrm{sink}$ and a
center of $\bmath{r}_i$, which coincides with the position of the $i$-th sink particle.
The sink particle obtains the mass
given by Equation~(\ref{eq:sink_mass_accretion}), and gas reduces the
density by $\Delta \rho(\bmath{r})$ 
according to Equation~(\ref{eq:sink_mass_accretion_density}).  
In the accretion process, the gas
velocity remains unchanged and the sink particle therefore obtains a linear
momentum of
\begin{equation}
\int_{|\bmath{r} - \bmath{r}_i| < r_\mathrm{sink}}\Delta \rho(\bmath{r}) \bmath{v}(\bmath{r}) dV.
\end{equation}
The accretion of the linear momentum changes the velocity of the sink
particle. The mass and the linear momentum are conserved between the gas
and the sink particles in the accretion process.

\subsection{Gravitational Interaction}

The sink particles are Lagrangian particles, moving according to 
the accretion of linear momentum described above and
the gravitational interaction.
The gravitational force of the $i$-th sink particle is 
\begin{equation}
\bmath{g}_{\mathrm{sink}, i}(\bmath{r}) = 
\left\{
\begin{array}{ll}
\displaystyle
-\frac{G m_i}{|\bmath{r}-\bmath{r}_i|^3}
(\bmath{r}-\bmath{r}_i) \;\; 
& \textrm{(for $|\bmath{r}-\bmath{r}_i| \geq r_\mathrm{soft} $)}\\
\displaystyle
-\frac{G m_i}{r_\mathrm{soft}^3}
(\bmath{r}-\bmath{r}_i) \;\; 
& \textrm{(for $|\bmath{r}-\bmath{r}_i| < r_\mathrm{soft} $)}
\end{array}
\right. , 
\label{eq:gravity_of_sink}
\end{equation}
where $m_i$ denotes the mass of the $i$-th sink particle.
Equation~(\ref{eq:gravity_of_sink}) represents the gravitational force of a
uniform sphere with a radius of $r_\mathrm{soft}$, which corresponds
to a softening radius and is set at $r_\mathrm{soft} = r_\mathrm{sink}$ here.
The gas is accelerated due to the self-gravity of the
gas and the sum of the gravitational forces of all the sink particles:
\begin{equation}
\bmath{g}(\bmath{r}) =
-\nabla \Psi(\bmath{r}) + 
\sum_i \bmath{g}_{\mathrm{sink}, i}(\bmath{r}).
\end{equation}
When computing the gravitational forces for a sink particle
acting on the cells within the softening radius ($|\bmath{r}-\bmath{r}_i| < r_\mathrm{soft} $), 
each cell is subdivided into $8^3$ subcells, and the gravitational
forces of the sink particle acting on the subcells are summed 
 \citep{Krumholz04}.

The gravitational force of gas acting on the sink particle is
evaluated as a reaction force of $\bmath{g}_{\mathrm{sink},i}$
\citep[see][]{Krumholz04}.  
This ensures conservation of
the linear momentum through the gravitational interaction between the
gas and the sink particle within a round-off error. 
An alternative method for evaluation of the gravitational force of gas
acting on the sink particle is taking
the average of the gravity of the gas over the spherical region 
with the softening radius: 
\begin{equation}
\bmath{g}_{\mathrm{gas}, i} = 
-
\frac{
\displaystyle
\int_{|\bmath{r} - \bmath{r}_i| < r_\mathrm{soft}}
\nabla \Psi (\bmath{r})
dV
}{
\displaystyle
\int_{|\bmath{r} - \bmath{r}_i| < r_\mathrm{soft}}
dV
}
\end{equation}
where each cell is subdivided into 
$8^3$ cells for integration. 
This alternative method is valid because Equation~(\ref{eq:gravity_of_sink})
coincides with the gravity of a uniform sphere. 
Finally, the gravitational force acting on the $i$-th sink particle is
given by
\begin{equation}
\bmath{g}_i = \bmath{g}_{\mathrm{gas},i} + \sum_{j\neq i} \bmath{g}_{\mathrm{sink},j}(\bmath{r}_i).
\end{equation}

\subsection{Time Integration}
The second-order leapfrog scheme is adopted for the time integration of
sink particles. 
Sink particles and hydrodynamics generally have a common
timestep.  The timestep $\Delta t$ is restricted so
that the travel distances of the particles in a step 
are less than the smallest cell width
of the numerical grid, 
\begin{equation}
\Delta t = \min(\Delta t_\mathrm{CFL}, \Delta t_\mathrm{vs})
\end{equation}
\begin{equation}
\Delta t_\mathrm{vs}  =  
C_\mathrm{vs} \min_{i}\left[ \frac{\Delta x }{ v_{i,x} } , \frac{\Delta y }{
    v_{i,y} } , \frac{\Delta z }{ v_{i,z} } \right],
\end{equation}
where $\Delta t_\mathrm{CFL}$ denotes a timestep determined by the 
Courant-Friedrichs-Lewy (CFL)
condition, and $\bmath{v}_i = (v_{i,x}, v_{i,y}, v_{i,z})$ denotes the
velocity of the $i$-th sink particle.
The constant $C_\mathrm{vs}$ is set at 0.5 here. 
When a sink particle is accelerated by strong gravity, 
the timestep is divided into a sub-timestep of 
\begin{equation}
\Delta t _\mathrm{sub} = 
C_\mathrm{gs}
\min_{i, j \ne i}
\left[
\left(
\frac{
\min(|\bmath{r}_i - \bmath{r}_j|, \Delta x, \Delta y, \Delta z)
}{
|\bmath{g}_i |
}
\right)^{1/2}
\right].
\end{equation}
The constant $C_\mathrm{gs}$ is set at 0.1 here. 
To reduce the computational cost, 
the gravity of gas acting on sink particles is approximated to
be unchanged during the sub-timesteps. 
Similar sub-cycling is also adopted in \citet{Krumholz04} and \citet{Federrath10a}.

\section{Decomposition of the velocity field into longitudinal and transverse components}
\label{sec:decomposition}

The velocity field $\bmath{v}(\bmath{r})$ can be separated into a
transverse (solenoidal) component $\bmath{v}_\perp(\bmath{r})$ and a longitudinal (compressive)
component $\bmath{v}_\parallel(\bmath{r})$
by applying the Helmholtz decomposition, where the decomposed velocity
fields are divergence-free ($\nabla \cdot \bmath{v}_\perp = 0$) and 
curl-free ($\nabla \times \bmath{v}_\parallel = 0$).
For the decomposition, a Fourier transform has been used 
\citep[e.g.,][]{Kitsionas09,Federrath10}.
The Fourier transform of the velocity fields,
\begin{eqnarray}
\tilde{\bmath{v}}_\perp(\bmath{k}) &=& \int \bmath{v}_\perp (\bmath{r}) \,e^{-i \bmath{k} \cdot \bmath{r}} \,d^3\bmath{r},\\
\tilde{\bmath{v}}_\parallel(\bmath{k}) &=& \int \bmath{v}_\parallel (\bmath{r}) \,e^{-i \bmath{k} \cdot \bmath{r}} \,d^3\bmath{r},
\end{eqnarray}
are evaluated as
\begin{eqnarray}
\tilde{\bmath{v}}_\parallel(\bmath{k}) &=& \hat{\bmath{k}}( \hat{\bmath{k}}\cdot\tilde{\bmath{v}}(\bmath{k})),\label{eq:vpark}\\
\tilde{\bmath{v}}_\perp(\bmath{k}) &=& \tilde{\bmath{v}}(\bmath{k}) - \tilde{\bmath{v}}_\parallel(\bmath{k}),\label{eq:vperpk}
\end{eqnarray}
with $\hat{\bmath{k}} = \bmath{k} / \left| \bmath{k} \right|$ and 
\begin{equation}
\tilde{\bmath{v}}(\bmath{k}) = \int \left(\bmath{v} (\bmath{r}) -
\langle\bmath{v}\rangle \right) \,e^{-i \bmath{k} \cdot \bmath{r}} \,d^3\bmath{r}.
\label{eq:fftv}
\end{equation}
In the integrand of Equation~(\ref{eq:fftv}), the volume-weighted mean velocity
$\langle \bmath{v} \rangle$ is subtracted from the original velocity
$\bmath{v}(\bmath{r})$ to obtain $\tilde{\bmath{v}}(0)=0$, 
because an orthogonality is not defined at $\bmath{k} =0$.
The transverse and longitudinal components are perpendicular 
($\tilde{\bmath{v}}_\perp \perp \bmath{k}$)
and
parallel 
($\tilde{\bmath{v}}_\parallel \parallel \bmath{k}$), respectively,
with respect to $\bmath{k}$.
The velocity spectra of the transverse and longitudinal modes are therefore given by
\begin{eqnarray}
E_\mathrm{trans}(k)\,dk &=& \frac{1}{2}\int_k^{k+dk} \tilde{\bmath{v}}_\perp\cdot\tilde{\bmath{v}}^*_\perp\,4\pi k^2dk, \\
E_\mathrm{long}(k)\,dk  &=& \frac{1}{2}\int_k^{k+dk} \tilde{\bmath{v}}_\parallel\cdot\tilde{\bmath{v}}^*_\parallel\,4\pi k^2dk.
\end{eqnarray}
By integrating $E_\mathrm{trans}(k)$ and $E_\mathrm{long}(k)$ over the whole $\bmath{k}$ space, 
the powers of the velocity components are obtained as 
\begin{eqnarray}
E_\mathrm{trans} &=&
\int E_\mathrm{trans}(k)\,dk = \frac{1}{2}\int \tilde{\bmath{v}}_\perp \cdot\tilde{\bmath{v}}^*_\perp \,d^3\bmath{k,} \label{eq:e_trans}\\
E_\mathrm{long} &=&
\int E_\mathrm{long}(k)\,dk  = \frac{1}{2}\int \tilde{\bmath{v}}_\parallel \cdot\tilde{\bmath{v}}^*_\parallel \,d^3\bmath{k}.\label{eq:e_long}
\end{eqnarray}
Note that Parseval's theorem states that 
\begin{eqnarray}
E_\mathrm{trans} &=& \frac{1}{2}\int \left|\bmath{v}_\perp\right|^2
\,d^3\bmath{r} = \frac{1}{2} \sigma^2_\perp L^3,\\
E_\mathrm{long}  &=& \frac{1}{2}\int \left|\bmath{v}_\parallel
\right|^2 \,d^3\bmath{r} = \frac{1}{2} \sigma^2_\parallel L^3,
\end{eqnarray}
where $\sigma_\perp$ and $\sigma_\parallel$ denote the velocity
dispersions due to the transverse and longitudinal velocity components, respectively.
The total velocity power coincides with 
the sum of the transverse and longitudinal powers because of the orthogonality
between $\tilde{\bmath{v}}_\perp$ and $\tilde{\bmath{v}}_\parallel$,
\begin{equation}
E_\mathrm{tot} = 
\frac{1}{2}\int \tilde{\bmath{v}}
\cdot\tilde{\bmath{v}}^*\,d^3\bmath{k}
= E_\mathrm{trans}+E_\mathrm{long}.
\end{equation}
This also indicates the additivity of the velocity dispersions,
$\sigma_v^2 = \sigma_\perp^2 + \sigma_\parallel^2$, where
$\sigma_v^2 = 2 E_\mathrm{tot}/L^3$.

\end{appendix}

\end{document}